\documentclass[aps,prd,nofootinbib,superscriptaddress,showpacs,preprintnumbers,twocolumn,12pts]{revtex4-1}
\usepackage{epsfig,dcolumn}
\usepackage{color,graphicx}
\usepackage{amsmath,amsfonts,amssymb}
\usepackage{bm} 
\usepackage{todonotes}
\usepackage{hyperref}
\hypersetup{ 
    pdfnewwindow=true,   
    colorlinks=true,     
    linkcolor=blue,      
    citecolor=blue,      
    filecolor=blue,      
    urlcolor=blue        
}
\usepackage{xspace}
\usepackage{slashed}
\usepackage{dsfont}

\newcommand{\diff}{\textrm{d}}

\newcommand{\elab}{\ensuremath{E_\text{lab}}\xspace}

\DeclareMathOperator{\im}{Im}

\newcommand{\ie}{{\it i.e.}\xspace}

\newcommand{\mev}{\ensuremath{{\mathrm{\,Me\kern -0.1em V}}}\xspace}
\newcommand{\gev}{\ensuremath{{\mathrm{\,Ge\kern -0.1em V}}}\xspace}
\newcommand{\tev}{\ensuremath{{\mathrm{\,Te\kern -0.1em V}}}\xspace}
\newcommand{\unitm}{\ensuremath{\mathds{1}}\xspace}

\usepackage{mfirstuc} 
\newcommand{\addReviewer}[2]{
  \expandafter\newcommand\csname #1\endcsname[1]{{\bf \color{#2} \capitalisewords{#1}:\,##1}}
  \expandafter\newcommand\csname #1cor\endcsname[2]{{\color{#2} \capitalisewords{#1}:\,\st{##1}{\bf ##2}}}
  \expandafter\newcommand\csname #1color\endcsname{#2}
}

\usepackage{tikz, marginnote} 
\newcommand{\checkedby}[1]{
\ifdefined\CROSSCHECKS
  \marginnote{
    \begin{tikzpicture}
      \foreach \x [count=\xi] in {#1} {
         \node[shape=circle,inner sep=0mm,
         minimum size=2mm,
         fill=\csname \x color\endcsname] at (\xi*3mm,0) {};
       }
    \end{tikzpicture}
  }
\else
\fi
}

\usepackage{soul,color}
\definecolor{chromeyellow}{rgb}{1.0, 0.65, 0.0}
\definecolor{DodgeBlue}{rgb}{0.118, 0.565,1.000}
\definecolor{asparagus}{rgb}{0.53, 0.66, 0.42}
\definecolor{cadmiumgreen}{rgb}{0.0, 0.42, 0.24}

\addReviewer{adam}{red}
\addReviewer{ale}{chromeyellow}
\addReviewer{vincent}{brown}
\addReviewer{misha}{cadmiumgreen}
\addReviewer{jannes}{blue}
\addReviewer{cesar}{magenta}
\addReviewer{tomasz}{cyan}
\addReviewer{miguel}{DodgeBlue}
\addReviewer{andrew}{purple}
\addReviewer{nathan}{orange}
\addReviewer{arkaitz}{asparagus}

\begin{document}

\title{Structure of Pion Photoproduction Amplitudes}

\author{V.~Mathieu}
\email{vmathieu@jlab.org}
\affiliation{Theory Center, Thomas Jefferson National Accelerator Facility,
12000 Jefferson Avenue,  Newport News, VA 23606, USA}
\author{J.~Nys}
\affiliation{Department of Physics and Astronomy, Ghent University, Belgium}
\author{C.~Fern\'andez-Ram\'irez}
\affiliation{Instituto de Ciencias Nucleares, Universidad Nacional Aut\'onoma de M\'exico, Ciudad de M\'exico 04510, Mexico}
\author{A.~N.~Hiller Blin}
\affiliation{Institut f\"ur Kernphysik \& PRISMA Cluster of Excellence, Johannes Gutenberg Universit\"at, D-55099 Mainz, Germany}
\author{A.~Jackura}
\affiliation{Center for Exploration of Energy and Matter, Indiana University, Bloomington, IN 47403, USA}
\affiliation{Physics Department, Indiana University, Bloomington, IN 47405, USA}
\author{A.~Pilloni}
\affiliation{Theory Center, Thomas Jefferson National Accelerator Facility,
12000 Jefferson Avenue,  Newport News, VA 23606, USA}
\author{A.~P.~Szczepaniak}
\affiliation{Center for Exploration of Energy and Matter, Indiana University, Bloomington, IN 47403, USA}
\affiliation{Physics Department, Indiana University, Bloomington, IN 47405, USA}
\affiliation{Theory Center, Thomas Jefferson National Accelerator Facility,
12000 Jefferson Avenue,  Newport News, VA 23606, USA}
\author{G. Fox}
\affiliation{School of Informatics and Computing, Indiana University, Bloomington, IN 47405, USA} 

\collaboration{Joint Physics Analysis Center}
\noaffiliation
\preprint{JLAB-THY-18-2755}

\begin{abstract}
We derive and apply the finite energy sum rules to pion photoproduction. We evaluate the low energy part of the sum rules using several state-of-the-art models. We show how the differences in the low energy side of the sum rules might originate from different quantum number assignments of baryon resonances. We interpret the observed features in the low energy side of the sum rules with the expectation from Regge theory. 
Finally, we present a model, in terms of a Regge-pole expansion, that matches the sum rules and the high-energy observables.
\end{abstract}

\maketitle

\section{Introduction}
Single pion photoproduction was the first measurement performed with the GlueX detector~\cite{AlGhoul:2017nbp} at the Jefferson Lab  and will likely 
 be one of the first measurements at CLAS12. 
   At low photon energies, $E_\gamma \sim \mathcal{O}(1\gev)$, it is a rich source of information on the baryon  spectrum~\cite{Drechsel:1998hk,FernandezRamirez:2005iv,Mariano:2007zza,Gasparyan:2010xz,Workman:2011vb,Anisovich:2011fc,Tiator:2011pw,Briscoe:2012ni,Kamano:2013iva,Ronchen:2014cna,Kamano:2016bgm},
 while at high energies, $E_\gamma \sim$ O(10 GeV), it reveals the details of hadron interactions 
 mediated by cross-channel particle (Reggeon) exchanges~\cite{Irving:1977ea}. These two energy regimes are analytically connected, a feature that can be used to relate the properties of resonances in the direct channel to the Reggeon exchanges in the crossed channels. In practice, this can be accomplished through dispersion relations and finite-energy sum rules (FESR)~\cite{Dolen:1967jr, Dolen:1967zz}.

There are several models in the literature focusing on neutral and charged pion photoproduction in the high energy region~\cite{Ader:1967tqj, Goldstein:1973xn,Goldstein:1973sf, Goldstein:1974mr, Guidal:1997hy, Sibirtsev:2007wk, Sibirtsev:2009kw,  Yu:2011zu, Mathieu:2015eia}. The differences between the various models are mainly  due to the fact that momentum transfer dependence of Regge pole residues is largely unknown. 
 In the past FESR were used to constrain 
  residues in either  neutral~\cite{Argyres:1974uk,Barker:1974vm, Barker:1977pm} or charged~\cite{Jackson:1969iv,Hontebeyrie:1973jx,Rahnama:1990mt} pion photoproduction independently,  
   and the fit to both reactions was performed by Worden in Ref.~\cite{Worden:1972dc}. 
Fixed-$t$ dispersion relations were also used in the past to determine the baryon spectrum in Refs.~\cite{Devenish:1974zh, Barbour:1974ie, Barbour:1976ne, Barbour:1978cx} but to the best of our knowledge, FESR in photoproduction  have not  been implemented in constraining the low-energy models. This is important because in the last decades high quality data in the low energy region have been collected and new partial wave analyses have been performed. These will be discussed in more detail later.  While the $N^*$ and $\Delta$ spectra below 2\gev are ``at least fairly well explored'' according to the PDG~\cite{pdg}, the properties of the higher excitations are poorly known. The 2-3\gev energy range is the transition between the baryon resonance region and the Regge regime. Since the number of relevant partial waves increases with energy, additional tools are required to constrain the amplitude construction. 
 As we show in this paper the analytical  constraints 
 from high energy can indeed be useful to improve the extraction of baryon resonances. 
This study complements our analysis of $\eta$ photoproduction~\cite{Nys:2016vjz} and 
pion-nucleon scattering~\cite{Mathieu:2015gxa}.  

The paper is organized as follows. In  Sec.~\ref{sec:formalism} we decompose the amplitudes into a covariant basis and define 
 the scalar amplitudes. The singularities of the latter are the only ones required by unitarity, which makes them suitable for a dispersive analysis. After reviewing the properties of the scalar amplitudes, we use dispersion relations and a standard Regge parametrization to derive the FESR in Sec.~\ref{sec:FESR}.  
In Sec.~\ref{sec:low} we evaluate the low-energy side of the sum rules with various available  partial wave models. We also extract the effective Regge residues and show 
   that the low-energy models provide a good,  qualitative prediction for the observables at high energies. 
 In Sec.~\ref{sec:fit} we present a combined fit of the parameters in the Regge expansion to both the FESR and the high-energy observables.  Our conclusions are presented in Sec.~\ref{sec:conclusion}.

\section{Formalism: Scalar Amplitudes} \label{sec:formalism}
The photoproduction of a pion off a nucleon (proton or neutron) target:
\begin{equation} \label{eq:reaction}
\gamma(k,\lambda_\gamma) + N(p,\lambda) \rightarrow \pi(q) + N'(p',\lambda')
\end{equation}
depends on three helicities ($\lambda_\gamma$, $\lambda$ and $\lambda'$) and the two Mandelstam variables: the center-of-mass energy squared $s=(k+p)^2$ and the momentum transferred squared $t =(q-k)^2$. The third Mandelstam variable $u = (p' - k)^2$ is fixed by the relation $s+t+u = 2m_N^2 + m_\pi^2$, where $m_x$  denotes the mass of the particle $x$. The helicities $\lambda_\gamma$, $\lambda$ and $\lambda'$ are defined in the center-of-mass of the reaction~\eqref{eq:reaction}, customarily denoted as the $s$ channel frame. The $t$ channel frame refers to the center-of-mass frame of the cross-channel reaction $\gamma \pi \to \bar N N'$. 

The photoproduction of a pseudoscalar is fully described by four scalar amplitudes. The standard Chew-Goldberger-Low-Nambu (CGLN) decomposition~\cite{Chew:1957tf} reads
 \begin{align}\label{eq:cgln_decomposition}
A_{\lambda'; \lambda \, \lambda_\gamma}(s,t) =
\sum\limits_{k = 1}^4  \overline{u}_{\lambda'}(p') A_k(s,t) M_k u_{\lambda}(p) \,.
\end{align}
The definition of the covariant basis $M_k\equiv M_k(s,t,\lambda_\gamma)$ and all relevant kinematical quantities can be found in Ref.~\cite{Mathieu:2015eia}.
In the following we neglect isospin violations. Writing explicitly the isospin indices ($i,j$ for the target and recoil nucleon respectively and $a$ for the isovector pion), the  $t$-channel isospin decomposition for each scalar amplitude $A_k$ (omitting the $k$ index) reads 
\begin{align}
A_{ji}^a & = A^{(0)}{\tau}^a_{ji} + A^{(+)} \delta^{a 3} \delta_{ji} 
+ A^{(-)}  \frac{1}{2} [\tau^a,\tau^3]_{ji} ,
\end{align}
with $\tau^a$ the Pauli isospin matrices. In this basis, $A^{(0)}$ is the amplitude involving the isoscalar component of the photon, while $A^{(+)}$ and $A^{(-)}$ involve the isovector one, with the $\gamma \pi$ system in isospin $0$ and $1$ respectively. 
More explicitly, the $t$ channel  
(\ie exchange) quantum numbers are
\begin{align}
I^G(A^{(0)} ) &= 1^+, &
I^G( A^{(+)} ) &= 0^-, &
I^G( A^{(-)} ) & = 1^-.
\end{align}
One could alternatively decompose into the $s$-channel isospin basis:
\begin{align} \nonumber
A_{ji}^a & = A^{(0)}{\tau}^a_{ji}  + A^{(1/2)}  \frac{1}{3}\left(\tau^a \tau^3 \right)_{ji} \\
& \qquad + A^{(3/2)} \left(\delta^{a3} \unitm -  \frac{1}{3} \tau^a \tau^3 \right)_{ji}.
\end{align}
In this basis, $A^{(0)}$ is the amplitude involving the isoscalar component of the photon, while $A^{(1/2)}$ and $A^{(3/2)}$ involve the isovector one, with the $\pi N$ system in isospin $1/2$ and $3/2$ respectively. 
\begin{subequations} \label{eq:schannelisospin}
\begin{align}
A^{(+)} & = \frac{1}{3} \left( A^{(1/2)} +2  A^{(3/2)} \right), \\
A^{(-)} & = \frac{1}{3} \left( A^{(1/2)}- A^{(3/2)} \right).
\end{align}
\end{subequations}
The isospin relations in Eq.~\eqref{eq:schannelisospin} suggest a connection between the baryon resonances, having definite $s$-channel quantum numbers, and the Regge exchanges with  definite $t$-channel quantum numbers.

The charged and neutral pion photoproduction reactions are described by an appropriate combination of the isospin components of the scalar amplitudes. Schematically, the contributions of isospin amplitudes to the helicity amplitudes are
\begin{subequations} \label{eq:reactions}
\begin{align}
A\left( \gamma p \to \pi^+ n \right)&= \sqrt{2} \left(A^{(0)} +A^{(-)} \right), \\
A\left(\gamma n \to \pi^- p\right)&=  \sqrt{2} \left(A^{(0)} -A^{(-)} \right),\\
A\left(\gamma p \to \pi^0 p \right)&= \phantom{\sqrt{2} \quad} A^{(+)} + A^{(0)}, \\
A\left(\gamma n \to \pi^0 n \right)&= \phantom{\sqrt{2} \quad} A^{(+)} - A^{(0)}.
\end{align}
\end{subequations}

The $u-$channel, $\gamma \bar N \to \pi \bar N$, is obtained from the $s$ channel by charge conjugation. Symmetry under charge conjugation implies definite parity for the scalar amplitudes under the transformation $s \leftrightarrow u$. This can be made explicit by using the symmetric variable
\begin{equation}
\nu = \frac{s-u}{4 m_N}  =   \elab + \frac{t-m_\pi^2}{4 m_N}\,,
\end{equation}
with $E_\text{lab}$ the photon energy in the laboratory frame (target rest frame). The scalar amplitudes can be separated into crossing-even
\begin{subequations} \label{eq:crossing}
\begin{align} \nonumber
A^{(0,+)}_{1,2,4}(-\nu-i \epsilon,t) & = + A^{(0,+)}_{1,2,4}(\nu+i \epsilon,t),\\
A^{(-)}_{3}(-\nu-i \epsilon,t) & = + A^{(-)}_{3}(\nu+i \epsilon,t),
\end{align}
and crossing-odd 
\begin{align} \nonumber
A^{(-)}_{1,2,4}(-\nu-i \epsilon,t) & = - A^{(-)}_{1,2,4}(\nu+i \epsilon,t),\\
A^{(0,+)}_{3}(-\nu-i \epsilon,t) & = - A^{(0,+)}_{3}(\nu+i \epsilon,t) ,
\end{align}
\end{subequations}
functions. 
In Refs.~\cite{Childers63,Mathieu:2015eia}, it was  shown that the scalar amplitudes 
$A_1$, $A_3$, and $A_4$ as well as the  $A_1+tA_2$ combination have also definite parity $P$ and naturality $P(-1)^J$ in the  $t$ channel. 
For convenience, we define
\begin{align}
A_2' \equiv A_1 + t A_2.
\end{align}
Table~\ref{tab:qn} summarizes the $t$-channel quantum numbers for the scalar amplitudes. 
In view of the symmetry relations in Eq.~\eqref{eq:crossing} we note that, with these standard conventions, 
the crossing-even (crossing-odd) amplitudes involve negative (positive) 
signature $\tau = (-1)^J$ exchanges.\footnote{The mismatch is simply coming from the extra $\nu$ factor in the $M_k$.} 
The exchanges are also divided into two other categories
according to {\it naturality}: 
the natural exchanges ($P(-1)^J=+1$) and the unnatural exchanges ($P(-1)^J=-1$). 
In addition to the signature and naturality of the exchanges we added in Table~\ref{tab:qn} the lowest spins and the name of the leading trajectory.\footnote{The leading or dominant trajectory is the Regge pole having the highest trajectory intercept $\alpha(0)$. Its contribution to the amplitude is thus the more important one, cf Eq.\eqref{eq:regge}.} 
We recall that the scalar mesons do not belong to the leading trajectories. The $\phi$ trajectory is also  sub-leading as its intercept is smaller. Moreover, the $\phi$ pole is expected to couple weakly to the nucleon. A recent estimation of the $\phi$ couplings to the nucleon can be found in Ref.~\cite{Mathieu:2017jjs}. 

\begin{table}[tbh]
\def\arraystretch{2.}
\centering
\caption{Invariant amplitudes $A_i$ with their corresponding $t$ channel exchanges.
$I$ is isospin, $G$ is $G$-parity, $J$ is total spin, $P$ is parity, $C$ is charge 
conjugation, and $\tau=(-1)^J $ is the signature. The name of the lightest meson on 
the trajectory is indicated in the last column. \label{tab:qn}}
\begin{ruledtabular}
\begin{tabular}{|>{$}c<{$}|>{$}c<{$}|>{$}c<{$}|>{$}c<{$}|>{$}c<{$}|>{$}c<{$}|}
A_i^{(\sigma)}	& \ I^G\  & P(-1)^J & \tau & J^{PC}	& \textrm{Lightest meson} \\
\hline \hline
A_{1,4}^{(0)} & 1^+ & +1 & -1	& (1,3,5,...)^{--}			& \rho(770)	\\
A_{1,4}^{(+)} & 0^- & +1 & -1 	& (1,3,5,...)^{--}			& \omega(782)	\\
A_{1,4}^{(-)} & 1^- & +1	& +1 & (2,4,6,...)^{++}			& a_2(1320)	\\
\hline
A_2'^{(0)} & 1^+  & -1	&-1	& (1,3,5,...)^{+-}		& b_1(1235)	\\
A_2'^{(+)} & 0^-  & -1	&-1	& (1,3,5,...)^{+-}		& h_1(1170)	\\
A_2'^{(-)} & 1^-  & -1	&+1	& (0,2,6,...)^{-+}		& \pi(140)	\\
\hline
A_3^{(0)} & 1^+ 	& -1	&+1	& (2,4,6,...)^{--}		& \rho_2(-)	\\
A_3^{(+)} & 0^-	& -1	&+1	& (2,4,6,...)^{--}		& \omega_2 (-)	\\
A_3^{(-)} & 1^-		& -1	&-1	& (1,2,5,...)^{++}		& a_1(1260)	\\
\end{tabular}
\end{ruledtabular}
\end{table}

Since crossed-channel exchanges control the behavior of the helicity amplitudes at high energy~\cite{Collins:1977jy, Irving:1977ea}, the $t$ channel quantum numbers of the scalar amplitudes are essential to determine their relative importance in the high energy region. Empirically, Regge trajectories involving natural exchanges dominate over unnatural trajectories.
Hence, from Table~\ref{tab:qn}, the scalar amplitudes $A_1$ and $A_4$ should contain the main contribution (\ie $\rho$, $\omega$, and $a_2$ exchanges) to the observables at high energies. We can obtain further indications of the high-energy behavior of the scalar amplitudes from their relation to the $s$-channel helicity amplitudes in the leading $s$ approximation
\begin{subequations} \label{def:Ai}
\begin{align}
\sqrt{-t} A_4 & = \frac{1}{\sqrt{2}s} \left( A_{+;+1}  + A_{-;-1}  \right), \\
\sqrt{-t} A_3 & = \frac{1}{\sqrt{2}s} \left( A_{+;+1}  - A_{-;-1}  \right), \\
 A_1 & = \frac{1}{\sqrt{2}s} \left( A_{+;-1}  - A_{-;+1}  \right), \\
A_2' & = \frac{-1}{\sqrt{2}s} \left( A_{+;-1}  + A_{-;+1}  \right),
\end{align}
\end{subequations}
where $\pm = \pm \frac{1}{2}$ is used for the nucleon helicities.
These relations show that, at the leading order in the energy, $A_3$ and $A_4$ are helicity non-flip at the nucleon vertex, and $A_1$ and $A_2'$ are helicity flip. 
It is well known that isoscalar  (isovector) exchanges are predominantly helicity non-flip (helicity flip) at the nucleon vertex~\cite{Irving:1977ea}. It is also known that the unnatural exchanges are suppressed at high energies, because of the smaller intercept.
Therefore, we expect $A^{(0,-)}_1$ and $A^{(+)}_4$
to dominate  at high energies.

Finally, the factorization of Regge pole residues yields a simple form for the kinematical singularities in $t$ at high energy~\cite{Cohen-Tannoudji:1968eoa}
\begin{align} \label{eq:tkin}
A_{\lambda'; \lambda \lambda_\gamma}(\nu,t) & \propto \left( \sqrt{-t} \right)^{|\lambda_\gamma|  + |\lambda'-\lambda|}.
\end{align}
From Eqs.~\eqref{def:Ai} and~\eqref{eq:tkin}, the Regge pole contributions in $A_1$ and $A'_2$ vanish in the forward direction, \ie $A_{1} \propto t$ and $A'_{2} \propto t$.
We now turn our attention to the analytic structure of the scalar amplitudes, 
and we derive the FESR in the next section.

\section{Finite Energy Sum Rules}\label{sec:FESR}
\begin{figure}[b]
\centerline{
\includegraphics[width=0.8\linewidth]{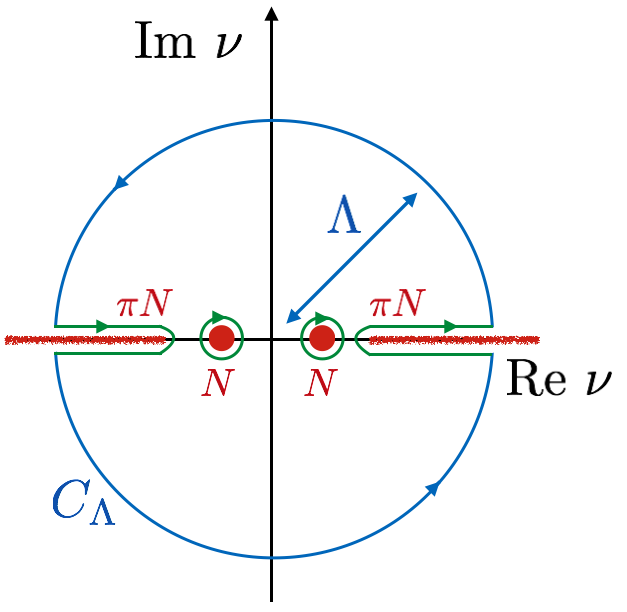}
}
\caption{\label{fig:DR} The complex $\nu$ plane. The singularities (nucleon pole and the two cuts starting at the $\pi N$ threshold) are in red. The integration contour is divided into two pieces as in Eq.~\eqref{eq:DRcut}, the contour surrounding the discontinuities and the circle $C_\Lambda$ of radius $\Lambda$. }
\end{figure}

The starting point of the FESR derivation is the analytic structure of the scalar amplitudes. The analytic structure and the associated dispersion relation for pion photoproduction are discussed extensively in the literature~\cite{Chew:1957tf, Ball:1961zza, Berends:1967vi,Pasquini:2006yi}. 
 The scalar functions have a nucleon pole and a left- and right-hand cuts required by unitarity, 
that are represented in the complex $\nu$ plane in Fig.~\ref{fig:DR}. 
The nucleon pole term is written, in our convention, as
\begin{align} \label{eq:pole}
\left. A_i^{(\sigma)}\right|_{\text{pole}} & = B_i^{(\sigma)}(t) \left( \frac{1}{\nu-\nu_N(t)} +\frac{\tau_i^{(\sigma)} }{\nu+\nu_N(t)} \right),
\end{align}
with $\nu_N(t) = (t-m_\pi^2)/(4m_N)$ the crossing variable at the nucleon pole. The nucleon pole residues $B_i^{(\sigma)}$ are tabulated in Table~\ref{tab:born}. According to Table~\ref{tab:qn}, the crossing-even (crossing-odd) scalar amplitudes correspond to Reggeons with negative (positive) signature $\tau_i^{(\sigma)}  = -1$ ($\tau^{(\sigma)}_i = +1$).

\begin{table}[htb]\caption{Residues of the Born term in eq.~\eqref{eq:pole} entering the dispersion relation. The pion pole in the residues $B_2^{(\sigma)}$ is canceled by a kinematic zero at $t=m_\pi^2$ in $M_2$. \label{tab:born}}
\begin{ruledtabular}
\begin{tabular}{| c | c | c |  c || c|}
 &&&& \\[-7pt]
$(\sigma)$ & $(0)$ & $(+)$ & $(-)$ &  \\[3pt]
\hline &&&& \\[-7pt]
$B_1^{(\sigma)}$ & $ -\frac{eg}{2m_N} \frac{1}{2}  $ &$-\frac{eg}{2m_N} \frac{1}{2} $ 
&  $-\frac{eg}{2m_N} \frac{1}{2} $ & $ e = 0.303 $ \\[5pt]
$B_2^{(\sigma)}$ & $ \frac{eg}{2m_N} \frac{1}{t-m_\pi^2 } $ &$ \frac{eg}{2m_N} \frac{1}{t-m_\pi^2 }$ &  $ \frac{eg}{2m_N} \frac{1}{t-m_\pi^2 }$ & $ g = 13.54 $  \\[5pt]
$B_3^{(\sigma)}$ & $ \frac{eg}{2m_N} \frac{\kappa_p+\kappa_n}{4m_N} $ 
& $ \frac{eg}{2m_N} \frac{\kappa_p-\kappa_n}{4m_N} $ 
& $ \frac{eg}{2m_N} \frac{\kappa_p-\kappa_n}{4m_N} $  & $ \kappa_p = 1.78 $  \\[5pt]
$B_4^{(\sigma)}$ & $ \frac{eg}{2m_N} \frac{\kappa_p+\kappa_n}{4m_N}  $ 
& $ \frac{eg}{2m_N} \frac{\kappa_p-\kappa_n}{4m_N} $ 
& $ \frac{eg}{2m_N} \frac{\kappa_p-\kappa_n}{4m_N}  $ & $ \kappa_n = -1.91$   \\[5pt]
\end{tabular}
\end{ruledtabular}
\end{table}

Let us consider the functions $\nu^k A_i(\nu,t)$ 
(we drop the isospin index) with $k$ being a positive integer. The functions $\nu^k A_i(\nu,t)$ have the same analytic structure as $A_i(\nu,t)$. Deriving the sum rules for $\nu^k A_i(\nu,t)$ instead of for $A_i(\nu,t)$ provides us with 
a set of constraints, or moments of order $k$. In Fig.~\ref{fig:DR}, we draw a contour in the complex $\nu$ plane. 
The contour surrounds  the singularities on the real axis (direct and cross-channel unitarity cuts and poles) and is closed with a circle of radius $\Lambda$. According to the Cauchy theorem the contour integral in Fig.~\ref{fig:DR} vanishes since analyticity requires the absence of singularities outside the real axis. Equivalently, we can match the discontinuity on the real axis to the integral along the circle of radius $\Lambda$
\begin{equation} 
\begin{split}
\int_{0}^{\Lambda} \left [ D_{i,R}(\nu,t) + (-1)^k D_{i,L}(\nu,t)  \right] \nu^k \frac{d \nu}{2i} \\  
= -\int_{C_\Lambda}  A_i (\nu,t) \nu^k \frac{d \nu}{2i},
\end{split}
\label{eq:DRcut}
\end{equation}
where
we include the nucleon poles in the discontinuities. For $\nu>0$, $D_{i,R}$ and $D_{i,L}$ correspond to the discontinuities along the $s$ channel (right) and $u$-channel (left) unitarity cuts respectively
\begin{subequations}
\begin{align}
D_{i,R}(\nu,t) = \lim_{\epsilon \to 0} 
\left[ A_i(+\nu+i \epsilon,t) - A_i(+\nu-i \epsilon,t) \right], \\
D_{i,L}(\nu,t) = \lim_{\epsilon \to 0} 
\left[ A_i(-\nu+i \epsilon,t) - A_i(-\nu-i \epsilon,t) \right].
\end{align}
\end{subequations}
Due to the crossing properties of the scalar functions, 
we can relate the left and right discontinuities 
$D_{i,L}(\nu,t) = \tau_i D_{i,R}(\nu,t)$. 
The left hand side (lhs) of the sum rule 
in Eq.~\eqref{eq:DRcut} becomes
\begin{align}
\left[1+ \tau_i(-1)^k \right] \int_{0}^{\Lambda} D_{i,R}(\nu,t) \nu^k \frac{d \nu}{2i}.
\end{align}
We note that the llhs of Eq.~\eqref{eq:DRcut} 
is nonzero only for $\tau_i=(-1)^k$ since $k$ is an integer. 
In other words, crossing-even (crossing-odd) amplitudes have odd (even) moments only. 

In our convention, the discontinuities include the nucleon pole at $\nu_N(t)$ and the unitarity cuts starting at 
$\nu_\pi(t)$, the $\pi N$ threshold, given by
\begin{align}
\nu_\pi(t) & = m_\pi + \frac{t+m_\pi^2}{4m_N}.
\end{align}
If $\nu_\pi(t)>0$, the left and right cuts do not overlap, and the amplitude is real in a part of the real axis. 
In this case the discontinuities along the cuts are given by the imaginary part of the amplitudes. 
The contribution of the right hand discontinuity to the sum rules reads
\begin{align} \nonumber
\int_{0}^{\Lambda}  D_{i,R}(\nu,t) \nu^k \frac{d \nu}{2i}  & = \pi B_i(t) \nu^k_N(t) \\
&+ \int_{\nu_\pi(t)}^\Lambda \im A_i(\nu,t) \nu^k d\nu.
\label{eq:integral17}
\end{align}
If $\nu_\pi(t)<0$, the left and right cuts overlap.
Nevertheless, one can still use a contour 
passing in between the two cuts and obtain the same dispersion relation as in Eq.~\eqref{eq:DRcut}.
The discontinuity is still given by the imaginary part 
of the amplitude
along the cut, since the function is analytic 
in $t$ and is real for $t>0$ along this cut.

To work out the right hand side (rhs) of Eq.~\eqref{eq:DRcut}, we assume that $\Lambda$ is large enough to approximate the amplitudes by a single Regge pole for each definite isospin scalar amplitude
along the circle
\begin{align} \label{eq:regge}
A_i(\nu,t) 
& = - \beta_i(t) \frac{\tau_i (r_i \nu)^{\alpha_i(t)}   + (-r_i \nu)^{\alpha_i(t)} }{ (r_i \nu)\sin\pi\alpha_i(t)},
\end{align}
where $\tau_i $, as for the lhs of Eq.~\eqref{eq:DRcut}, 
is the signature of the exchange. $\beta_i(t)$ and $\alpha_i(t)$ are the residue and the trajectory of the Regge pole, respectively. 
The $r_i>0$ are scale factors required by dimensional analysis. They are of the same order of the typical hadronic scale in the process, O(1 \mbox{ GeV}). A change in the scale factor $r_i$ amounts simply to re-scaling the residue by an exponential factor.   
The $\nu$ factor in the denominator is meant to cancel the factor of $1/s$ in Eq.~\eqref{def:Ai}, stemming from the kinematic terms in Eq.~\eqref{eq:cgln_decomposition}, to provide the correct behavior $s^{\alpha_i(t)}$ of the helicity amplitudes in the large $s$ limit. On the real axis, Eq.~\eqref{eq:regge} reduces to the well-known form
\begin{align}
A_i(\nu,t) & = - \beta_i(t) \frac{\tau_i  + e^{-i\pi \alpha_i(t)} }{\sin\pi\alpha_i(t)} (r_i \nu)^{\alpha_i(t)-1}.
\end{align}

Assuming the form in Eq.~\eqref{eq:regge}, the integral along the circle of radius $\Lambda$ can be calculated analytically. 
The integration is performed separately for the two terms in Eq.~\eqref{eq:regge} as they have different cuts, 
\ie
a left hand cut for the first term and a right hand cut for the second. 
The first term contribution to the contour integral in Eq.~\eqref{eq:DRcut}, with the change of variable 
$\nu = \Lambda e^{i\phi}$, is
\begin{equation}
\begin{split}
 \tau_i  \beta_i(t) \frac{(r_i \Lambda)^{\alpha_i(t)-1} }{2i \sin\pi\alpha_i(t)} \Lambda^{k+1} 
\int_{-\pi}^{\pi} e^{i\phi(\alpha_i(t)+k)} i \: d\phi \\
 =
\tau_i(-1)^k \beta_i(t) \frac{(r_i \Lambda)^{\alpha_i(t)-1}}{\alpha_i(t)+k} \Lambda^{k+1}.
 \end{split}\label{eq:piece1}
\end{equation}
The second term yields the contribution to the contour integral
\begin{equation} 
\begin{split}
-\beta_i(t)\frac{(-r_i \Lambda)^{\alpha_i(t)-1}}{2i \sin\pi\alpha_i(t)}  \Lambda^{k+1}  \int_{0}^{2\pi} e^{i\phi(\alpha_i(t)+k)} id\phi \\
 = \beta_i(t)\frac{(r_i \Lambda)^{\alpha_i(t)-1}}{\alpha_i(t)+k} \Lambda^{k+1}.
\end{split} \label{eq:piece2}
\end{equation}
As expected, the rhs of Eq.~\eqref{eq:DRcut} 
also vanishes unless $\tau_i = (-1)^k$. 
Hence, we can combine Eqs.~\eqref{eq:integral17},
~\eqref{eq:piece1} and~\eqref{eq:piece2}
to obtain the FESR
\begin{equation}\label{eq:FESR}
\begin{split}
\pi B_i(t) \frac{\nu_N^k(t)}{\Lambda^{k+1}}
+ \int_{\nu_\pi(t)}^\Lambda \im A_i(\nu,t) \frac{\nu^k d\nu}{\Lambda^{k+1}} \\
\qquad = \beta_i(t)  \frac{(r_i \Lambda)^{\alpha_i(t)-1}}{\alpha_i(t)+k}.
\end{split}
\end{equation}
It should be kept in mind that the FESR 
in Eq.~\eqref{eq:FESR} are valid only 
for odd (even) values of $k$ for crossing-even (crossing-odd) amplitudes. It seems at first that the high energy side of the FESR has a pole at $\alpha(t)=0$ for $k = 0$. This situation may happen in the physical region for the leading trajectory, cf.~\eqref{eq:alpha}. In this case, the ghost pole at $\alpha(t) = 0$ in even-signature amplitudes forces a zero in the residue, \ie $\beta(t) \propto \alpha(t)$, making the rhs of Eq.~\eqref{eq:FESR} finite. We will check this prediction in the Section where we will evaluate the rhs of the FESR. In order to explicitly see the zeros in the residues we will always choose $k\ge 1$. 
In our derivation, we explicitly assumed a single Regge pole for each definite isospin scalar amplitude. 
In general, the rhs of the FESR involves as many terms as there are Reggeons or Regge cuts contributing to the amplitude.

The FESR in Eq.~\eqref{eq:FESR} was derived using the known analytic structure of the scalar amplitudes at fixed $t\le 0$. 
For large negative values of $t$, singularities coming from two fixed poles appear,
\ie box diagrams with internal pions and nucleons. 
They manifest as an additional cut parallel to the unitarity cut. Nevertheless, they are far away from the forward angle region. 
The closest singularity of the double spectral representation 
is at $t = -1.1\gev^2$ and $s>(1.6\gev)^2$, as shown in Ref.~\cite{Noelle:1977jq}. 
In this work, we focus on the forward region 
$-1\le t/\text{GeV}^2 \le 0$, hence we do not need 
to consider any additional singularity. 

\section{The low energy side of the sum rules} \label{sec:low}
\subsection{The models}
\begin{figure}[htb]
\centerline{
\includegraphics[width=0.9\linewidth]{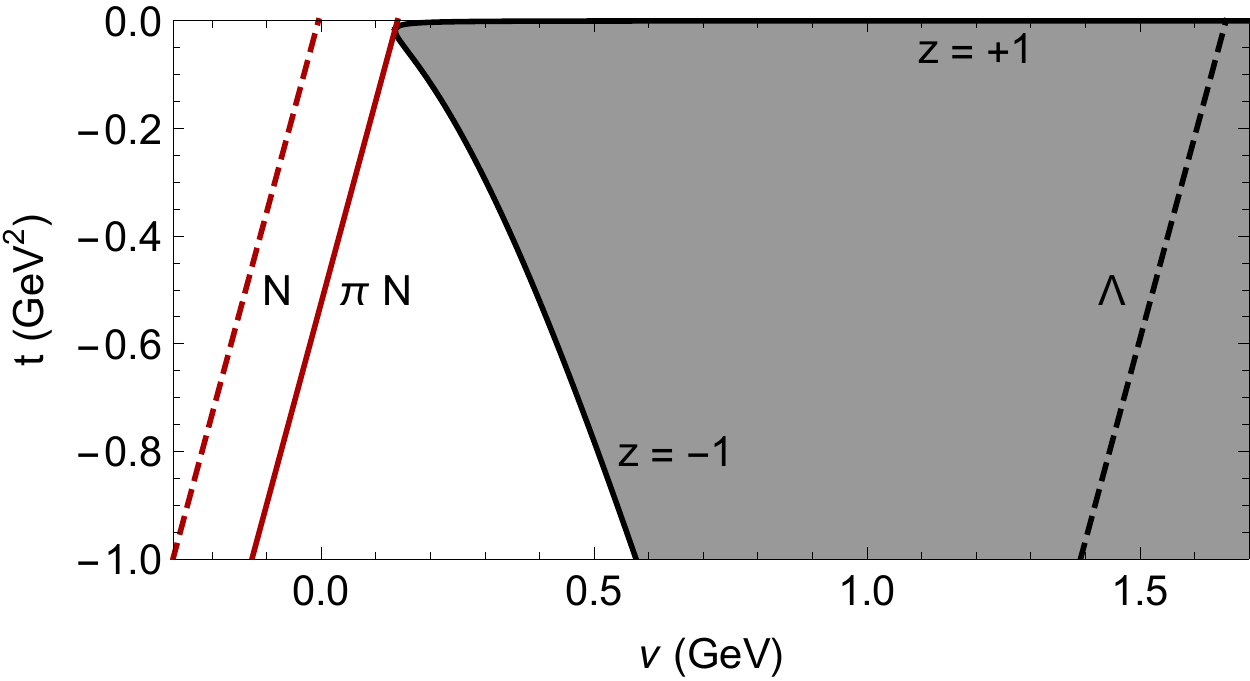}
}
\caption{\label{fig:nutplane}Low-energy region under investigation in this work in the $\nu-t$ plane. For fixed value of $t$, the integration region in $\nu$ for the lhs of the FESR is indicated by the red solid line (the $\pi N$ threshold) and the black dashed line (the cutoff $\Lambda$). The physical region of the process $\gamma N \to \pi N$ is indicated by the gray shaded area, limited by $z \equiv \cos\theta = \pm 1$.}
\end{figure}

\begin{figure*}[htb]
\centerline{
\includegraphics[width=\linewidth]{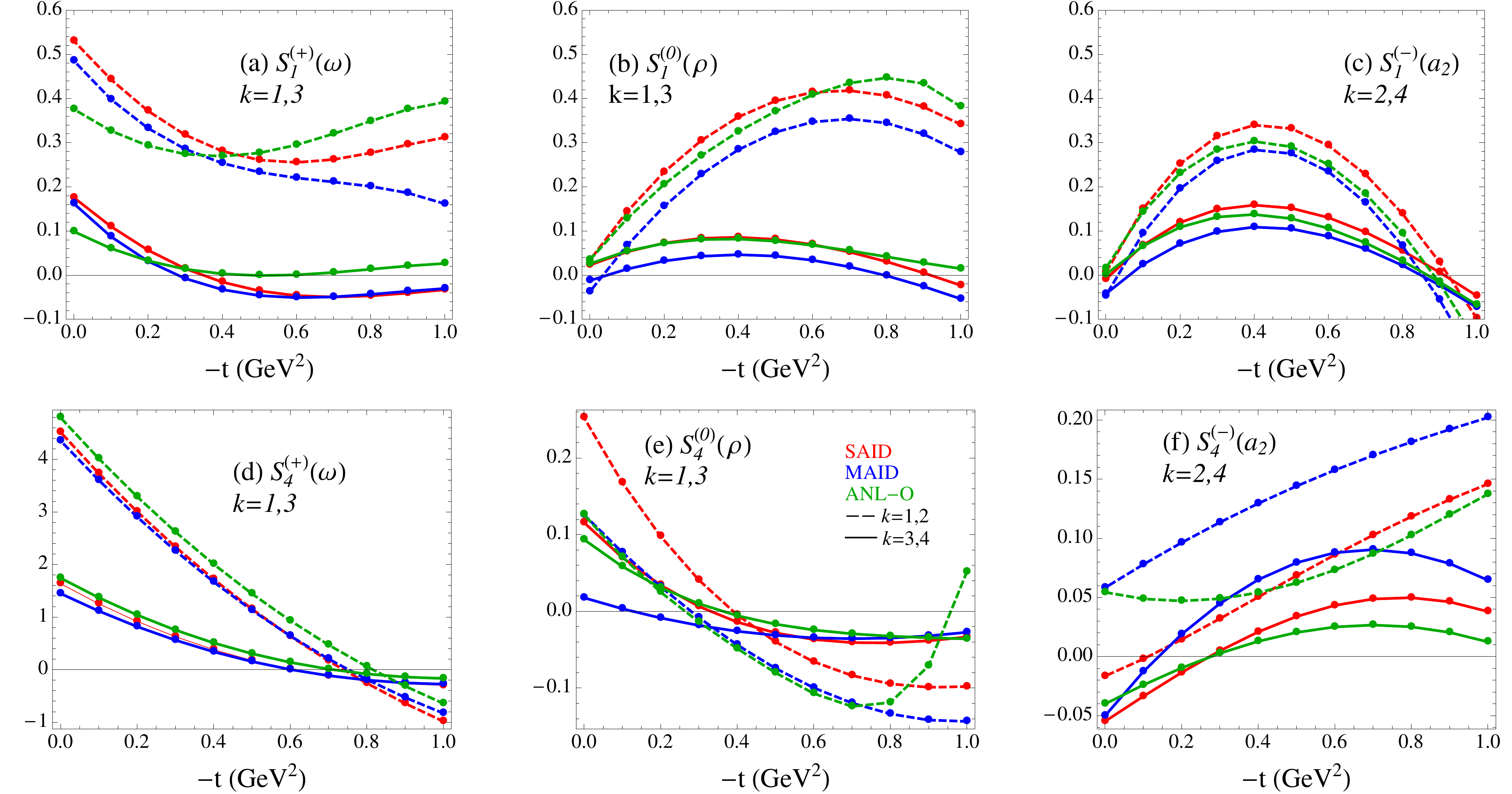}
}
\caption{\label{fig:plotFESR14a} First moments of the rhs of the FESR Eq.~\eqref{eq:rhs} for $A_{1,4}^{(0,\pm)}$ with SAID (red), MAID (blue) and ANL-O (green) models. The lowest spin particle on the corresponding Regge trajectory is indicated for convenience. The dashed (solid) lines correspond to the $k=1$ or $k=2$ ($k=3$ or $k=4$) moments and the cutoff is $s_\text{max} = 4$ GeV$^2$.}
\end{figure*}

There are several independent analyses of the baryon spectrum from photoproduction data. In this work, we will reconstruct the low energy side of the FESR using the five main amplitude models,  MAID with the MAID2007 version \cite{Drechsel:2007if}, SAID with the CM12 version \cite{Workman:2012jf}, Bonn-Gatchina (BnGa) with the BG2016 version \cite{CLAS-inprep},
J\"ulich-Bonn (J\"uBo) with the J\"uBo2014 version \cite{Ronchen:2014cna}, and ANL-Osaka (ANL-O) with the  ANL-O2016 version \cite{Kamano:2013iva,Kamano:2016bgm}.
The different models are compared in Ref.~\cite{Beck:2016hcy}. 
In this Section we first review the domain of validity of each model, and then evaluate the rhs of the sum rules in Eq.~\eqref{eq:FESR}  using the latest partial waves analysis by the different groups. 

The SAID, MAID and ANL-O groups include pion photoproduction on both a proton and a neutron target in their analyses while the latest J\"uBo and BnGa models are developed for proton targets only. Consequently SAID, MAID and ANL-O scalar amplitudes $A_i^{(\sigma)}$ are available for all isospin configurations, $\sigma = 0,+,-$, while for J\"uBo and BnGa we can  analyze $\gamma p \to \pi^0 p$ only. Indeed the left-hand-cut discontinuity of $\gamma p \to \pi^+ n$ is related to the physical region of the reaction $\gamma n \to \pi^- p$ by charge conjugation. Hence, the analysis of charged pion photoproduction requires both $\gamma n \to \pi^- p$ and $\gamma p \to \pi^+ n$ in the physical region.

The energy range of the different models and the number of multipoles available  are ($L$ being the angular momentum between the pion and the nucleon)
\begin{subequations} \nonumber
\begin{align}
\text{SAID:}   && \sqrt{s} \le 2.40 \text{ GeV and} \: L \le 5, \\
\text{MAID:}   && \sqrt{s} \le 2.00 \text{ GeV and} \, L \le 5, \\
\text{ANL-O:}  && \sqrt{s}\le 2.10 \text{ GeV and}  \: L \le 5 \\
\text{J\"uBo:} && \sqrt{s} \le 2.57 \text{ GeV and} \: L \le 5, \\
\text{BnGa:}  && \sqrt{s} \le 2.50 \text{ GeV and}  \: L \le 9.
\end{align}
\end{subequations}
The formulas to reconstruct the amplitudes from the multipoles are given in the Appendix F in Ref.~\cite{Nys:2016vjz}. 
We evaluate the lhs of the sum rule at fixed $t$ defined by
\begin{align}
S^{(\sigma)}_i(t,k) & \equiv \pi B^{(\sigma)}_i\frac{\nu^k_N}{\Lambda^{k+1}} + \int_{\nu_\pi(t)}^{\Lambda} \im A^{(\sigma)}_i(\nu,t)  \frac{\nu^k \diff \nu}{\Lambda^{k+1}},
\label{eq:rhs}
\end{align}
at 11 equally spaced points in the range $t \in [-1,0]\gev^2$. In the rest of the paper, we will discuss the sum rules computed with the amplitude $A_2^{\prime(\sigma)} = A_1^{(\sigma)} + t A_2^{(\sigma)}$. In order to simplify the notation, we will denote this quantity by $S^{(\sigma)}_2$.
In Eq.~\eqref{eq:rhs}, the dependence of the Born term on $t$ is understood, \ie $B_i^{(\sigma)} \nu_N^k(t) \equiv B_i^{(\sigma)}(t) \nu^k_N(t)$. The integral cutoff in $\nu$ can be made $t-$dependent, by expressing it in terms of a cutoff in energy  $s_\text{max}$:
\begin{align}\label{eq:cutoff}
\Lambda \equiv \Lambda(t) & = \frac{s_\text{max} -m_N^2}{2 m_N} + \frac{t-m_\pi^2}{4 m_N}.
\end{align}
The region of integration is indicated in Fig.~\ref{fig:nutplane}.
In the area outside the physical region, the amplitudes need to be extrapolated. 
In the unphysical region the cosine of the scattering angle reaches unphysical values $\cos\theta < -1$,  
but because the low energy models are reconstructed from multipoles, the $\cos\theta$ dependence is polynomial and given explicitly by Legendre polynomials. 
For high angular momenta in the multipole expansion, 
numerical instabilities could appear as the expansion goes as $(\cos \theta)^{L_\text{max}}$.
We have checked that the scalar functions, reconstructed with the five models, are continuous in the unphysical region if we use a partial waves expansion truncated up to $L_\text{max}=5$. Only the BnGa model has higher spin multipoles. For consistency with all other models, we truncate it  to $L_\text{max}=5$ as well.

\subsection{The low energy side for all isospin components}
The quantity in Eq.~\eqref{eq:rhs} computed with SAID, MAID and ANL-O models is presented in Figs.~\ref{fig:plotFESR14a} and \ref{fig:plotFESR23a} for all isospin components and the first two moments ($k=1,3$ for the crossing-even amplitudes and $k=2,4$ for the crossing-odd amplitudes). 
We choose $s_\text{max} = 4$ GeV$^2$, which is the highest energy  all the models can be pushed to.

\begin{figure}[h]
\centerline{
\includegraphics[width=0.7\linewidth]{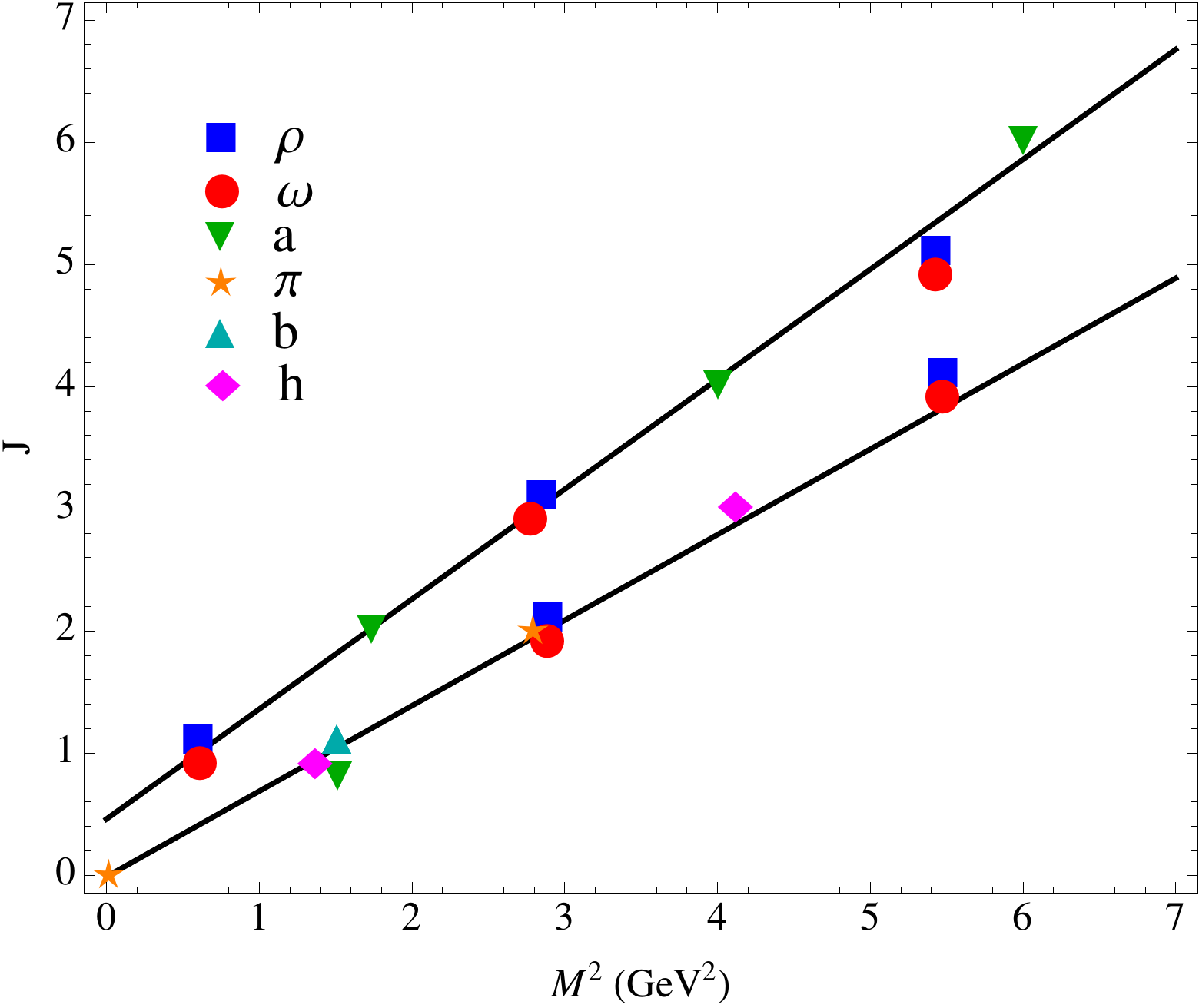}
}
\caption{\label{fig:trajec} Chew-Frautschi plot for natural and unnatural parity mesons. The solid lines indicate the two Regge trajectories $\alpha_N$ and $\alpha_U$ in Eq.~\eqref{eq:alpha}. The meson masses are taken from the Review of Particle Properties~\cite{Patrignani:2016xqp} except for the $2^{--}$ $\rho_2$ and $\omega_2$ mesons taken from a quark model calculation~\cite{Godfrey:1985xj}. }
\end{figure}

Since the factor $\Lambda^{\alpha_i(t)-1}/(\alpha_i(t)+k)$ in Eq.~\eqref{eq:FESR} never vanishes, zeros of $S^{(\sigma)}_i(t,k)$ should indicate the position of zeros in the ($k$-independent) residues $\beta_i(t)$. The moment independence of the zeros in $S^{(\sigma)}_i(t,k)$ is a good confirmation of the single Regge pole approximation. We then study the $S^{(\sigma)}_i(t,k)$ quantities given by low energy models and we 
compare them to the expectations from Regge theory. Since the position of zeros in the residues, and thus in the low energy side of the FESR Eq.~\eqref{eq:rhs}, can be related to the Regge trajectories, it is useful to have them in mind. The leading trajectory of each amplitude is fairly well known~\cite{Irving:1977ea}: 
\begin{subequations} \label{eq:alpha}
\begin{align}
\alpha_{1,4}^{(\sigma)} \equiv \alpha_N(t ) &= 0.9(t-m_\rho^2) + 1& &\text{for all }\sigma, \\
\alpha_{2,3}^{(\sigma)} \equiv \alpha_U(t ) &= 0.7(t-m_\pi^2) + 0 &  &\text{for all }\sigma.
\end{align}
\end{subequations}
These can be compared to meson masses for $t>0$ in Fig.~\ref{fig:trajec}.

\begin{figure*}[htb]
\centerline{
\includegraphics[width=\linewidth]{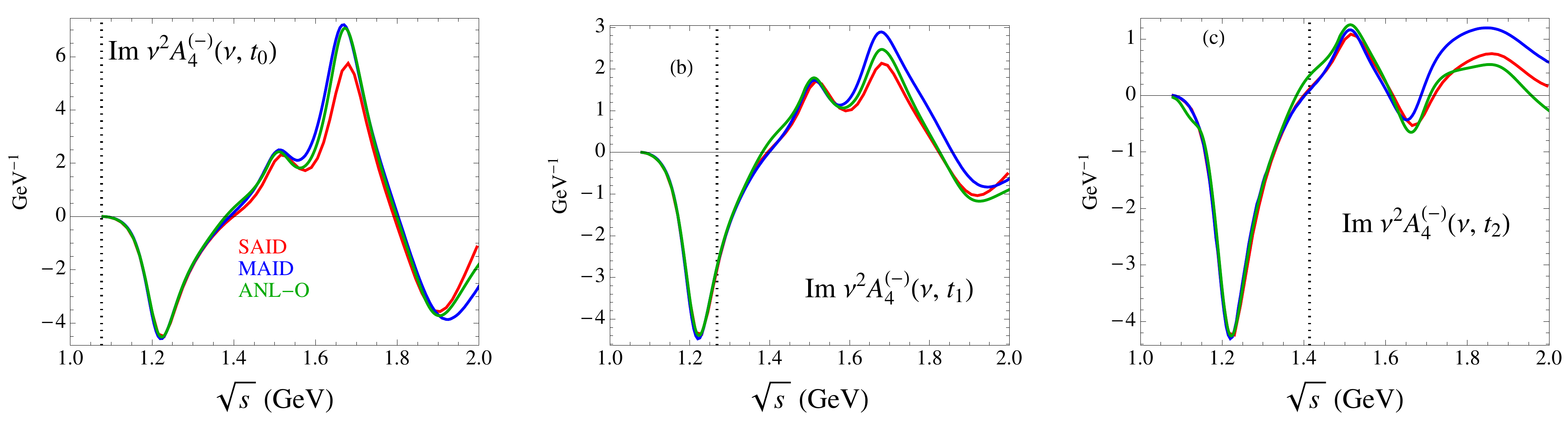}}
\caption{\label{fig:plotA4m} The imaginary part of the SAID (red), MAID (blue) and ANL-O (green) invariant amplitudes $\nu^2 A_4^{(-)}$ at $t_0=0$ GeV$^2$, $t_1=-0.3$ GeV$^2$ and $t_2=-0.6$ GeV$^2$. 
The vertical dashed line displays the beginning of the physical region.}
\end{figure*}

\begin{figure*}[htb]
\centerline{
\includegraphics[width=\linewidth]{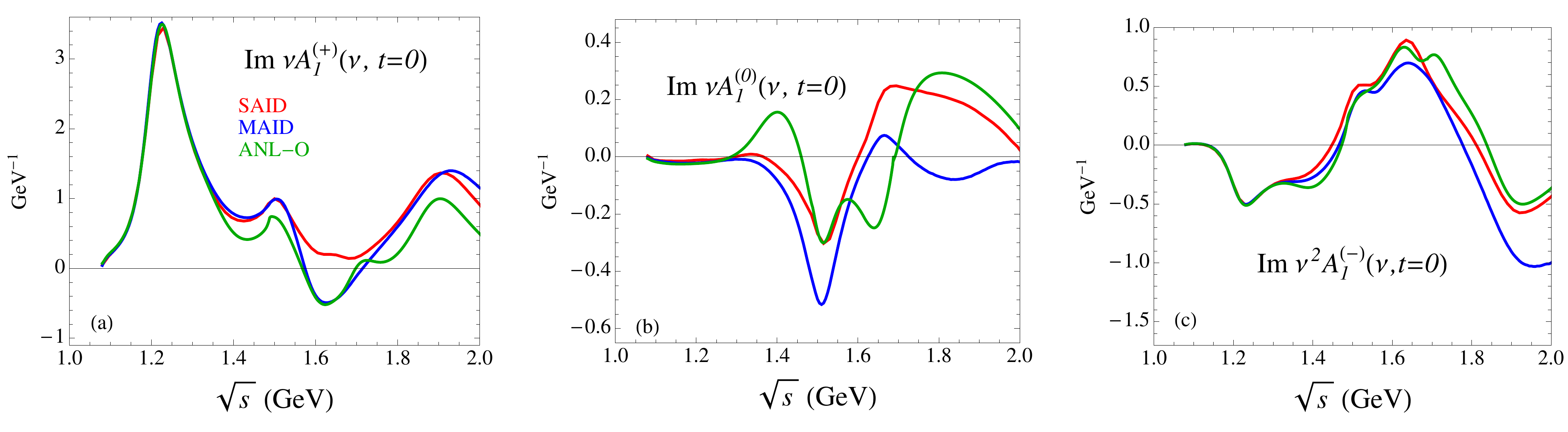}
}
\caption{\label{fig:plotA1t0} The imaginary part of the SAID (red), MAID (blue) and ANL-O (green) invariant amplitudes 
$\nu A_1^{(0,\pm)}$ at $t=0$. 
The $\Delta (1232)$ resonance is responsible 
for peaks at 1.2 GeV in $A_1^{(\pm)}$
and the non-vanishing $S_1^{(+)}(t=0,k)$ integral.
As expected from isospin symmetry 
$\Delta$ resonances do not contribute to $A_i^{(0)}$.}
\end{figure*}

\begin{figure*}[htb]
\centerline{
\includegraphics[width=\linewidth]{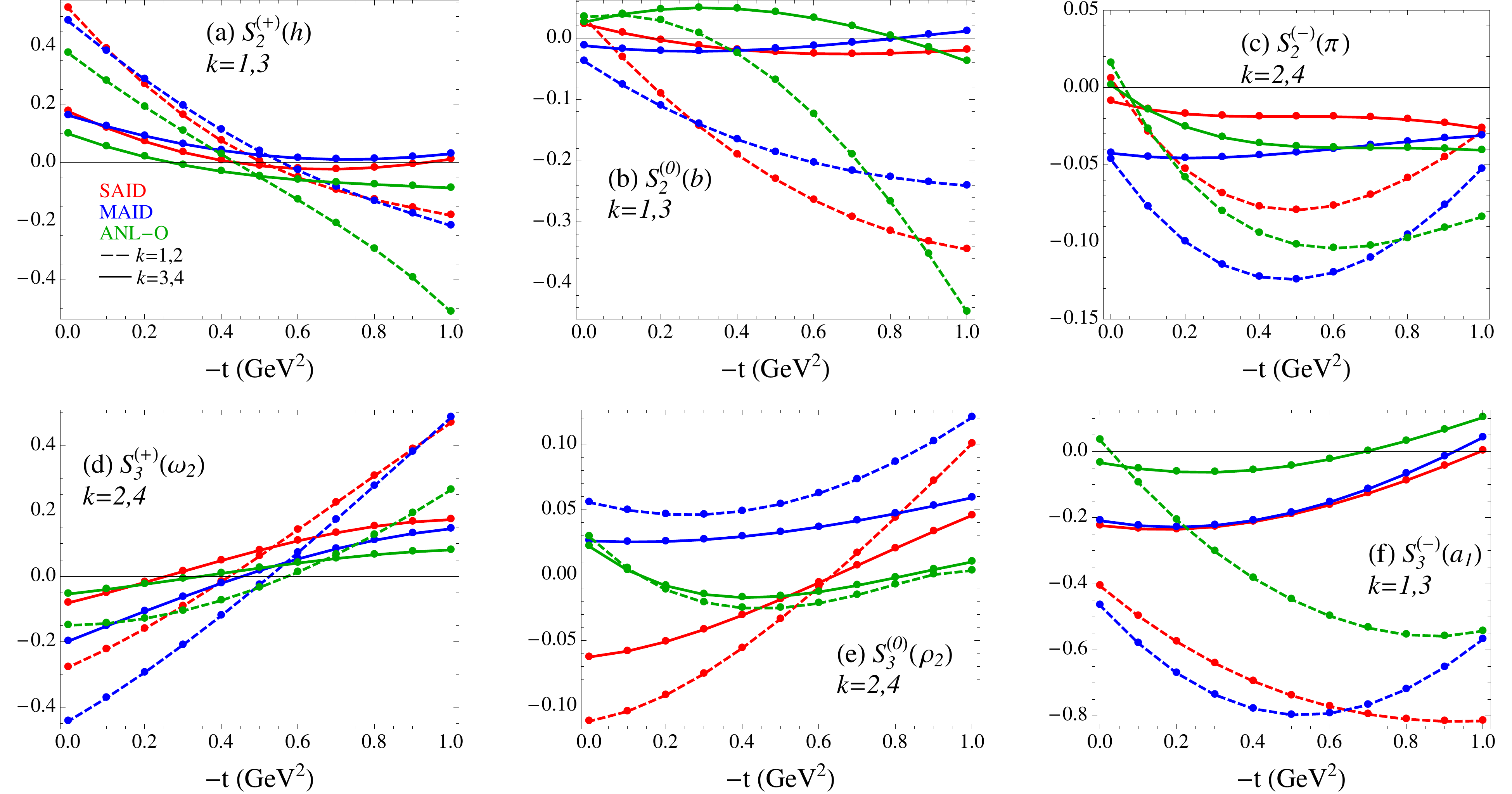}
}
\caption{\label{fig:plotFESR23a} First moments of the rhs of the FESR Eq.~\eqref{eq:rhs} for $A_{2,3}^{(0,+,-)}$ with SAID (red), MAID (blue) and ANL-O (green) models. The lowest spin particle on the corresponding Regge trajectory is indicated for convenience. The dashed (solid) lines correspond to the moment $k=1$ or $k=2$ ($k=3$ or $k=4$) and the cutoff is $s_\text{max} = 4$ GeV$^2$.}
\end{figure*}

\begin{figure*}[htb]
\centerline{
\includegraphics[width=\linewidth]{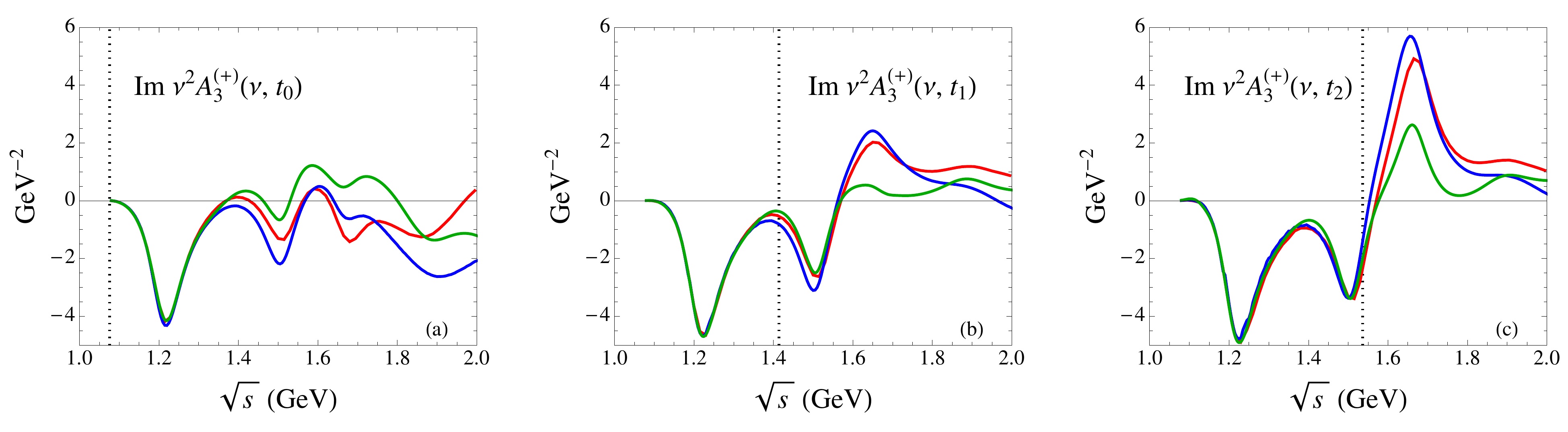}}
\caption{\label{fig:plotA3p} The imaginary part of the SAID (red), MAID (blue) and ANL-O (green) invariant amplitudes $\nu^2 A_3^{(+)}$ at $t_0=0$ GeV$^2$, $t_1=-0.6$ GeV$^2$ and $t_2=-0.9$ GeV$^2$. 
The vertical dashed line displays the beginning of the physical region. The magnitude of the Born term is represented by the horizontal dot-dashed line.}
\end{figure*}
In Fig.~\ref{fig:plotFESR14a}, we present the low energy side of the sum rules for the natural exchanges $S^{(\sigma)}_{1,4}(t,k)$, computed for the first two moments $1 \le k \le 4$ using the SAID, MAID and ANL-O models. We observe the following features:
\begin{enumerate}
\item All the three models shown give qualitatively similar results for all natural exchange $S^{(\sigma)}_{1,4}$'s. The strongest deviation between the model is observed in $S_4^{(-)}$. The imaginary part of the amplitude $\nu^2 A_4^{(-)}$ shown in Fig.~\ref{fig:plotA4m} does not vary drastically between the models. Nevertheless, the cancellation between the $\Delta(1232)$ and the other resonances results in a small $S_4^{(-)}$. The small differences in the structures at $\sqrt{s} = 1.7$~GeV and $\sqrt{s} = 1.9$~GeV are therefore magnified by the FESR. 
The deviation between the models at $\sqrt{s} = 1.7$~GeV can be traced back from the different magnitudes of the $N(1675)5/2^-$ and $N(1680)5/2^+$ resonances.
\item All the $S^{(\sigma)}_{i}$'s exhibit a zero in the range $t\in [-1,0]$ GeV$^2$, with the exception of the lowest moments $S_1^{(0,+)}(t,k=1)$ and $S_4^{(-)}(t,k=2)$. We identify two types of zeros. The ones at $t \sim -0.8\gev^2$ in $S^{(0,+)}_4$ and $S^{(-)}_1$ look independent of the moment, and most certainly correspond to  zeros in their corresponding Regge residues. Conversely, the ones at $t \sim -0.3\gev^2$ in $S^{(+)}_1$ and $S^{(0,-)}_4$ do not appear in the lowest moment. A possible reason may be the presence of sub-leading Regge contributions (daughter  trajectories and/or Regge cuts), whose importance decreases in higher moments. One can indeed check that the relative importance of a sub-leading trajectory $\alpha_2$ compared to the leading trajectory $\alpha_1 > \alpha_2$ is proportional to $(\alpha_1+k)/(\alpha_2+k)$, which decreases with $k$.   
\item The natural explanation for the zeros in $S^{(-)}_4(t,k)$ is the unwanted pole at $\alpha(t\sim -0.5\gev^2) = 0$. This pole would appear at a negative mass squared and must be canceled by a zero in the residue. Such a zero is called a nonsense wrong signature zero (NWSZ)~\cite{Collins:1977jy}. However, the zero in $S^{(0,-)}_1(t,k)$ is at $t \sim -0.8\gev^2$, significantly away from the expected position.
This zero might be shifted by the addition of another contribution 
(a daughter trajectory or a Regge cut) in the sum rules. 
A nonlinear trajectory with a zero at $t\sim -0.8$ GeV$^2$ 
would also explain this observation. A NWSZ should also appear in $S^{(-)}_1(t,k)$ for the same reason. The position of the zeros in  $S^{(-)}_1(t,k)$ and $S^{(-)}_4(t,k)$ would be at the same place with only one Regge pole contributing to the $A_{1,4}^{(-)}$ amplitudes. But the zero in $S_1^{(-)}$ appears at $t \sim -0.3\gev$ and another zero possibly arises at $-t > 1 \gev$.  We thus conclude that non leading trajectories are present in the $A_{1,4}^{(-)}$ amplitudes.
\item The position of the zeros in $S_{1,4}^{(0)}$ are very similar to the ones in $S_{1,4}^{(-)}$. Their origin can be explained by invoking the degeneracy between 
the $\rho$ and $a_2$ nucleon couplings, which is related to the absence of exotic resonances in $pp$ scattering~\cite{Mandula:1970wz,Irving:1977ea}. 
\item Inspecting the behavior in the forward direction, we see differences in the isoscalar  $S_1^{(+)}(0,k)$ and the isovector $S_1^{(0,-)}(0,k)$. The latter vanishes $\propto t$, while the former is finite. Also, the former is strongly $k$ dependent.
We have already observed such a pattern in $\eta$ photoproduction~\cite{Nys:2016vjz}. In pion photoproduction,
this effect is due to the contribution of the $\Delta(1232)$ resonance to $A^{(+)}_1$.
In Fig.~\ref{fig:plotA1t0} 
we show $\im \nu A^{(0,+)}_1$ and $\im \nu^2 A^{(-)}_1$ 
at $t=0$.  We observe that both $A^{(+)}_1$ and $A^{(-)}_1$
have a peak at $\sqrt{s}=1.2\gev$,  due to the $\Delta(1232)$. We can indeed check that at the peak $A^{(+)}_1 \approx -2A^{(-)}_1$ and $A^{(0)}\approx 0$, in agreement with Eq.~\eqref{eq:schannelisospin}, and with the dominance of a $I=3/2$ resonance.\footnote{At the $\Delta(1232)$ peak in the forward direction, $\nu\sim 0.33$ GeV.} 
In the isovector exchange amplitudes $A_1^{(0)}$ and $A_1^{(-)}$ 
the contributions of baryon resonances  cancel out to yield $S_1^{(0,-)}(t=0,k)\approx 0$. However, in $A_1^{(+)}$ the  contribution of the $\Delta$ is not canceled completely by other resonances, and produces a finite $S_1^{(+)}(t=0,k)$. This is in contrast with the factorization of Regge pole residues.
\item Among all of the natural exchange amplitudes, $S_4^{(+)}$ is one order of magnitude larger than the other ones. This effect can also be traced back to the fact that the dominant $\Delta(1232)$ contributes mainly to the isoscalar exchange amplitude. This is also consistent with the well-known dominance of the $\omega$ Regge pole in pion photoproduction. The non-flip nucleon couplings of isoscalar trajectories are larger than the ones for isovector exchanges~\cite{Irving:1977ea}. Moreover, in photoproduction there is an additional relative factor of 3 at the photon vertex between isoscalar and isovector exchanges.  All three models, SAID, MAID and ANL-O, provide very similar results for this dominant amplitude.
\item Interestingly, $S_4^{(+)}(t,k)$ has a zero at large $|t|$. The zero is around $t=-0.75\gev^2$ for the lower moment, and moves to $t=-0.6\gev^2$ for $k=3$. In the leading Regge pole  approximation, this zeros of the dominant Regge residue would imply a dip in the differential cross section at high energy in neutral pion photoproduction. This dip is indeed present at high energy, as shown in Fig.~\ref{fig:fitObs}. It is usually interpreted as NWSZ, although it is not mandatory in odd signature Regge poles, \ie there is no unphysical pole at $\alpha=0$ due to the signature factor.
\end{enumerate}

\begin{figure*}[htb]
\centerline{
\includegraphics[width=\linewidth]{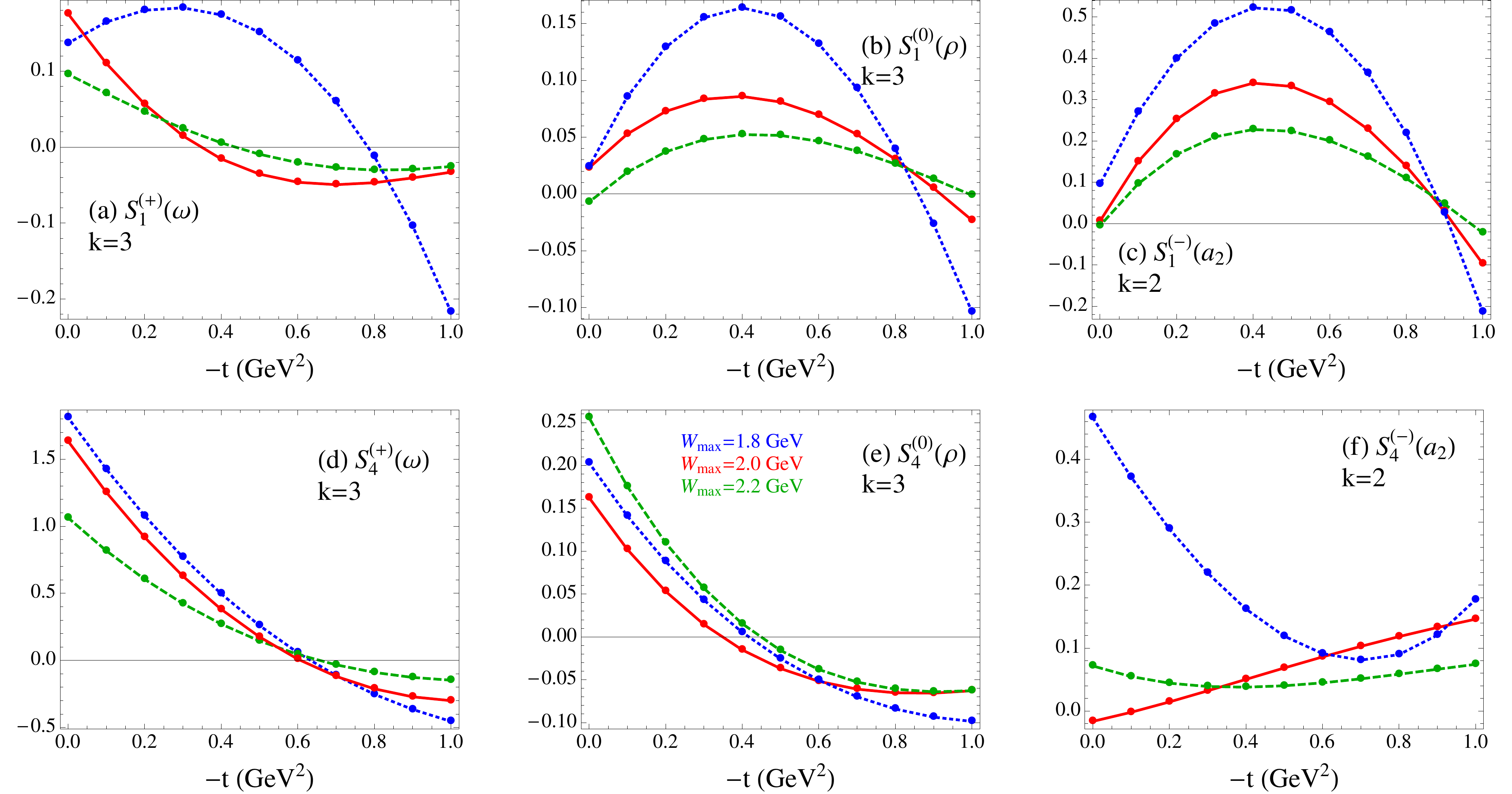}
}
\caption{\label{fig:cut14} First moments of the rhs of the FESR Eq.~\eqref{eq:rhs} for $A_{1,4}^{(0,\pm)}$ with SAID for three cutoffs: $s_\text{max} = (1.8\gev)^2$ (blue), $(2.0\gev)^2$ (red) and  $(2.2\gev)^2$ (green).}
\end{figure*}
\begin{figure*}[htb]
\centerline{
\includegraphics[width=\linewidth]{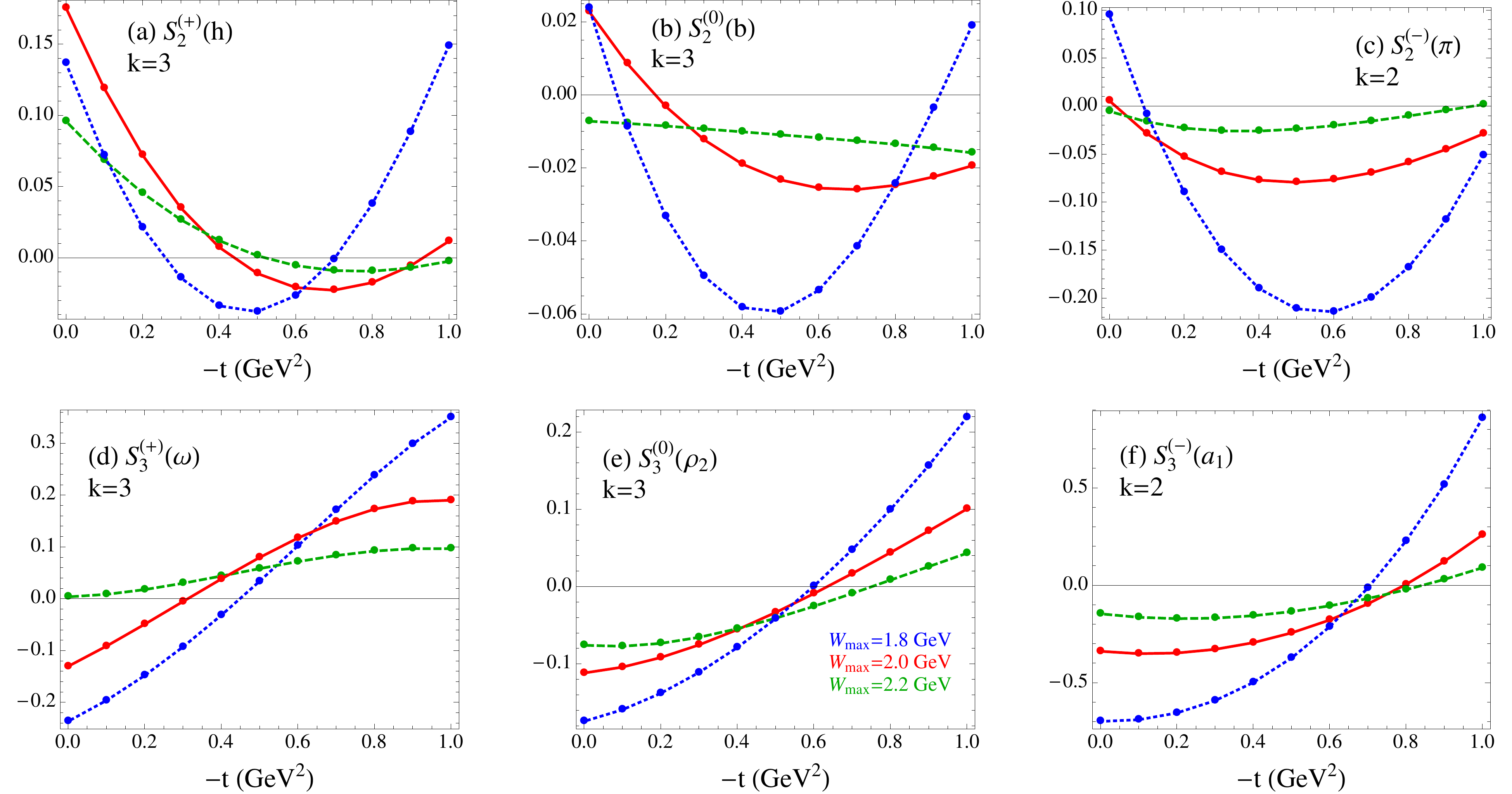}
}
\caption{\label{fig:cut23} First moments of the rhs of the FESR Eq.~\eqref{eq:rhs} for $A_{2,3}^{(0,\pm)}$ with SAID for three cutoffs: $s_\text{max} = (1.8\gev)^2$ (blue), $(2.0\gev)^2$ (red) and  $(2.2\gev)^2$ (green).}
\end{figure*}

In Fig.~\ref{fig:plotFESR23a}, we show the low energy side of the sum rules, for the unnatural exchanges, $S^{(\sigma)}_{2,3}(t,k)$. We compute those for the first two moments $k = 2,4$ using the SAID, MAID and ANL-O models and observe the following features. 
\begin{enumerate}
\item The difference between the three models for unnatural exchange amplitudes is significantly larger than for natural exchange amplitudes. This happens because of the large cancellation among the various resonant contributions, which makes the $S_{2,3}^{(\sigma)}$ particularly sensitive to the details of the resonance lineshapes. 
\item For the unnatural exchanges, there is no clear pattern of the zeros. The only exceptions are $S_{2,3}^{(+)}$, which both show a zero for $t \approx -0.5 \gev^2$.
\item The unnatural exchange terms are of the same order of magnitude as the natural ones, with the exception of the larger $S_4^{(+)}$ discussed above.
\item The factorization of Regge residues appears to be satisfied reasonably in $S^{(0)}_{2}$. However, $S^{(+)}_{1}$ and $S^{(+)}_{2}$ deviate significantly from the expected $\propto t$ behavior. Since $A_2' = A_1 + t A_2$, the deviation from the factorisable behavior in $S_2$ originates from the $A_1$ amplitude. The $\Delta(1232)$ peak in $A_1^{(0)}$ leads then to 
 a finite $S^{(+)}_{2}$ at $t=0$.  
\item The moment $S_2^{(-)}$ from the MAID models favors a nonzero value at $t=0$. The SAID and ANL-O models favor vanishing residues in the forward direction or possibly a zero at $|t|<0.1\gev^2$.  In charged pion photoproduction, the forward peak in the differential cross section requires a finite residue at $t=0$ in the pion exchange amplitude $A_2^{(-)}$. The beam asymmetry in charged pion photoproduction requires a zero at $t\sim 0.03\gev^2$ in the same amplitude. Both requirements are met with the $k=2$ moment with the SAID and ANL-O models. 
\item The exchanges $\omega_2$ and $\rho_2$ contributing to the amplitudes $A_3^{(0,+)}$ are poorly known and generally assumed to be small. This is consistent with the high energy data, as we will see, that does not favor a large $A_3$ contribution. This is in contrast with the sizable  $S_3^{(0,+)}$. The monotonic grow of $S_3^{(+)}$ can be deduced from Fig.~\ref{fig:plotA3p}. 
The $\Delta(1232)$ ($J^P = 3/2^+$), the $N(1520)$ ($3/2^-$) and the $N(1520)$ ($1/2^-$) have a mild $t$-dependence, but the higher-spin $N(1680)$ ($5/2^+$) contribution grows as $|t|$ increases, yielding a $S_3^{(+)}$ growing with $|t|$. Similar conclusions can be obtained for the $A_3^{(0)}$ amplitude. In order to obtain a negligible residue for all $t$ in the $A_3^{(0,+)}$ amplitudes, as the high energy data suggest, one would need to change the $t$ dependence. For instance a change in the spin-parity assignment of the $N(1680)$, currently $5/2^+$, to $3/2^+$ could result in a small value of  $S_3^{(+,0)}$.  
\end{enumerate}

From all these observations we conclude that the FESR amplifies the differences between various models. Due to the cancellation among resonances, the relative importance of higher-mass resonances is stronger in the FESR than in the original amplitudes. Moreover, FESR relate the $t$ dependence of the Regge residues to the spin of the $N^*$ and $\Delta$ resonances. In general we notice that moments  $k=2$ and $k=3$  are in the best agreement with the expectation from Regge theory and in the following focus on these moments.

Although we explained that the FESR can point out the differences between models, by looking only at one side of the sum rule we cannot claim whether one model is better or worse than another. Secondly we do not have information concerning the uncertainties associated to these multipoles. These uncertainties, propagated through the scalar amplitudes and then in the FESR, would certainly provide useful information. At this stage, we cannot conclude if the observed differences in the sum rules between the various models  are coming from the model dependence, or rather by the data uncertainties in the low energy region. The amplitudes of the available models are not fully constrained, since a complete set of observables is not yet available~\cite{Chiang:1996em,Sandorfi:2010uv,Nys:2016uel,Landay:2016cjw}. On can expect that 
 if double polarization measurements were included  the low-energy models could change as shown for example in Ref.~\cite{Strauch:2015zob}.

\subsection{Cutoff dependence}
With the multipoles provided by the SAID group, we can investigate the dependence of the cutoff $s_\text{max}$ in the sum rules. In Fig.~\ref{fig:cut14} and~\ref{fig:cut23}, we plot the low energy side of the FESR $S^{(\sigma)}_i(t,k)$ for $s_\text{max} = (1.8\gev)^2$, $(2.0\gev)^2$ and  $(2.2\gev)^2$. We observe that the positions of the zeros in the natural exchange amplitudes, $S_{1,4}^{(\sigma)}$, are relatively stable when the cutoff is varied. The notable exceptions are $S_1^{(+)}$ and $S_4^{(-)}$, when evaluated at $s_\text{max} = (1.8 \gev)^2$. The amplitudes $A_1^{(+)}$ and  $A_4^{(-)}$indeed receive a significant contribution from the $\Delta(1930)$ resonance as can be seen in Figs.~\ref{fig:plotA4m} and~\ref{fig:plotA1t0}. Some moments, {\it e.g.} the $S_{3}^{(0)}$  have significant $s_\text{max}$ dependence. 
A possible explanation is that the underlying 
 amplitudes are less constrained, or that they are more sensitive to higher mass resonances. In other words, the uncertainties associated to some of the curves in Figs.~\ref{fig:cut14} and~\ref{fig:cut23} could  be significant. 

In the following we choose an ``optimal'' cutoff. As we saw, $s_\text{max} = (1.8\gev)^2$ is too low. Since we do not observe a drastic change between $s_\text{max} = (2.0\gev)^2$ and $s_\text{max} = (2.2\gev)^2$, we will choose $s_\text{max} = (2.0\gev)^2$. With all models being valid at least up to that energy, we will be able to compare their moments.

\subsection{The low energy side for $\gamma p \to \pi^0 p$}
\begin{figure*}[t]
\centerline{
\includegraphics[width=\linewidth]{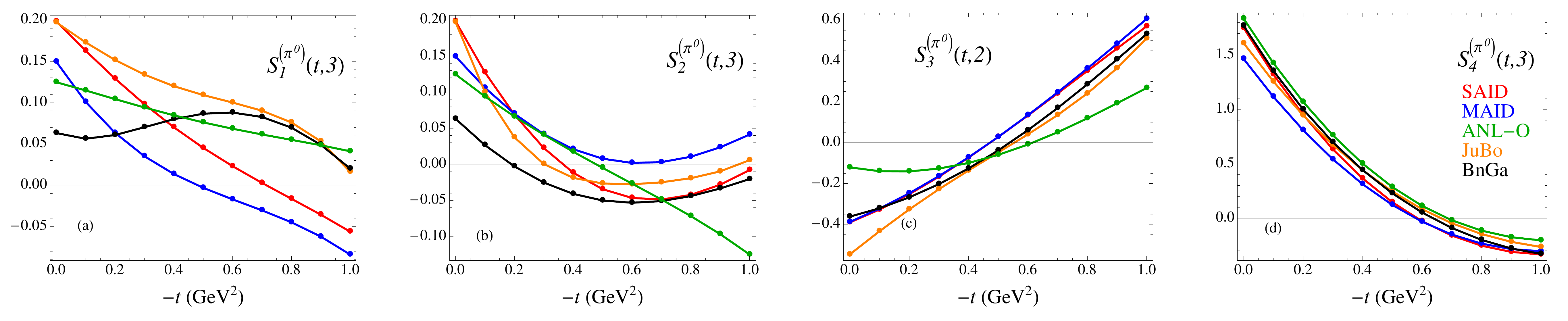}
}
\caption{\label{fig:plotFESRJuBo} Lowest moments of the rhs of the FESR Eq.~\eqref{eq:rhs} for $A_{i}^{(\pi^0)} = A_{i}^{(0)} + A_{i}^{(+)}$ with SAID, MAID and J\"uBo ($L_\text{max}=5$ is used) solutions for the process $\gamma p \to \pi^0 p$. The integral is truncated at $s_\text{max} = (2.0\gev)^2$.}
\end{figure*}

\begin{figure*}[t]
\centerline{
\includegraphics[width=\linewidth]{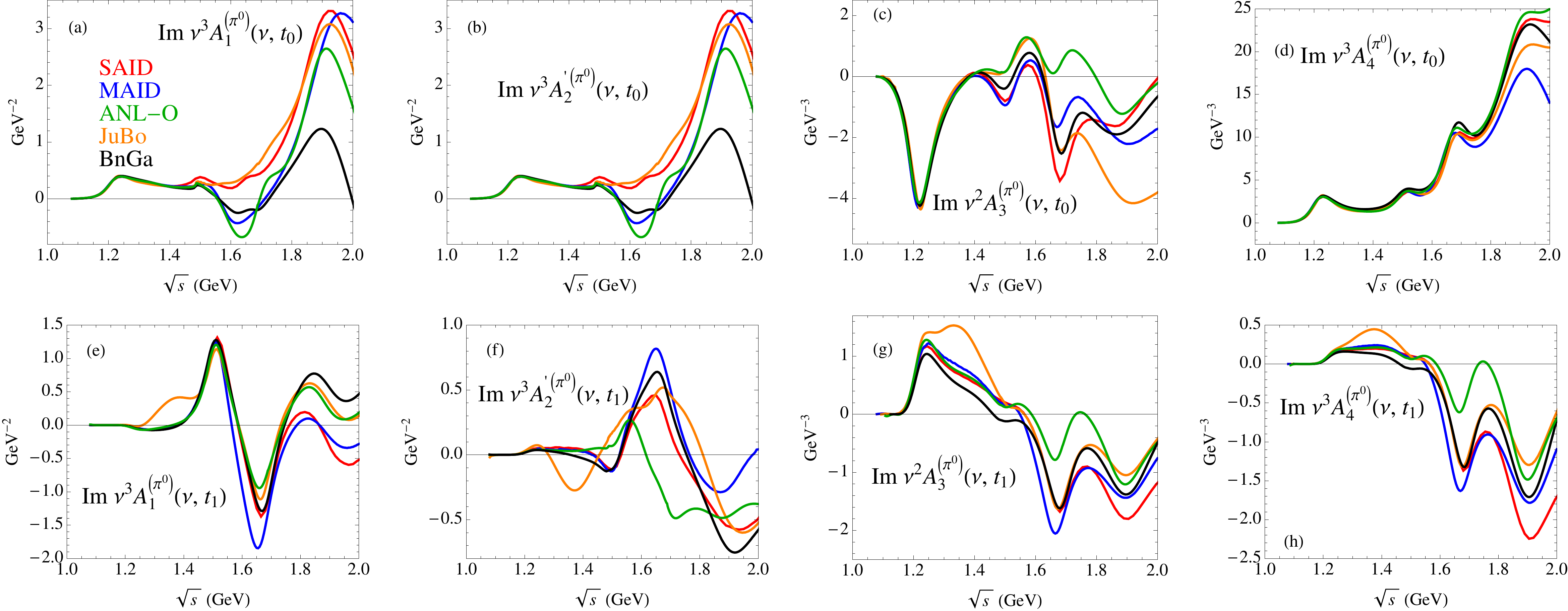}
}
\caption{\label{fig:plotAiAll} The invariant amplitudes $A_i$ with SAID, MAID, ANL-O, BnGa and J\"uBo models for the process $\gamma p \to \pi^0 p$ at $t_0=0$ and $t_1=-0.8\gev^2$.}
\end{figure*}

For completeness, we compare the FESR obtained with the J\"uBo and BnGa models with the SAID, MAID and ANL-O models. The J\"uBo and BnGa models are only available for reactions on a proton target. As stated, we can only present the results for the process $\gamma p \to \pi^0 p$, beacuse for $\gamma p \to \pi^+ n$ the FESR require the knowledge of $\gamma n \to \pi^- p$ to evaluate the left-hand cut.

The comparison between J\"uBo, BnGa, SAID, MAID and ANL-O models is shown  in Fig.~\ref{fig:plotFESRJuBo}. The cutoff $s_\text{max}=(2.0 \text{ GeV})^2$ is used in the FESR and only the moment $k=2$ or $k=3$ is plotted. The J\"uBo and BnGa models compare very well with the SAID, MAID and ANL-O models except for $S_1^{(\pi^0)} = S_1^{(0)} + S_1^{(+)}$. We can identify the cause of this difference by looking at the invariant amplitudes at fixed $t$. 
We compare in Fig.~\ref{fig:plotAiAll}, the four scalar functions for the neutral pion photoproduction reconstructed from the SAID, MAID, ANL-O, BnGa and J\"uBo multipoles, as a function of the energy at $t_0 = 0$ and $t_1 = -0.8$ GeV$^2$. We note that all models yield similar scalar amplitudes up to $\sqrt{s} \sim 1.6\gev$, but the relative strengths and $t$-dependence of the resonances beyond this region differ. The higher moments give stronger weight to the heavier resonances, and thus amplify the differences between the various models. Imposing the FESR constraints in the PWA analyses will certainly reduce the variation between them, and yield more accurate $N^*$ and $\Delta$ spectra.

\subsection{$t$-channel amplitudes}
\begin{figure*}[htb]
\centerline{
	\includegraphics[width=0.49\linewidth]{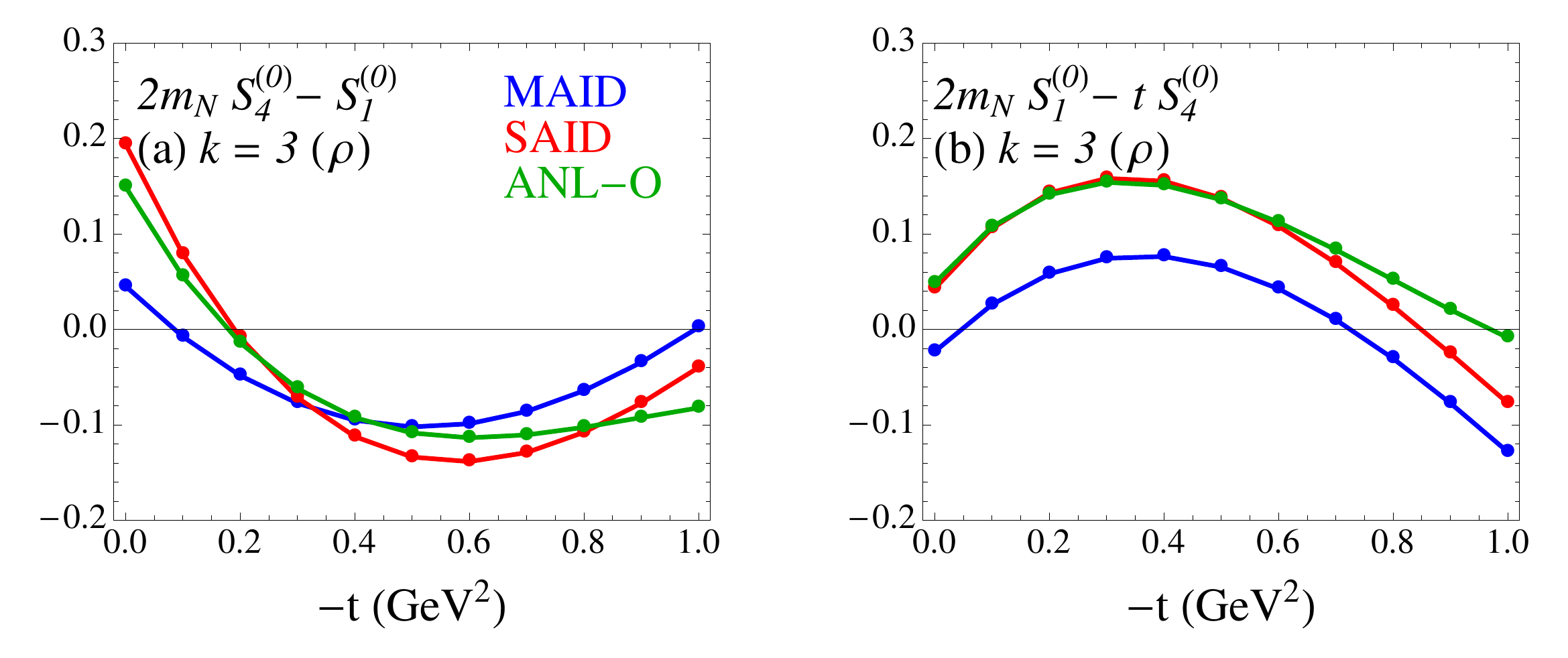}
	\includegraphics[width=0.49\linewidth]{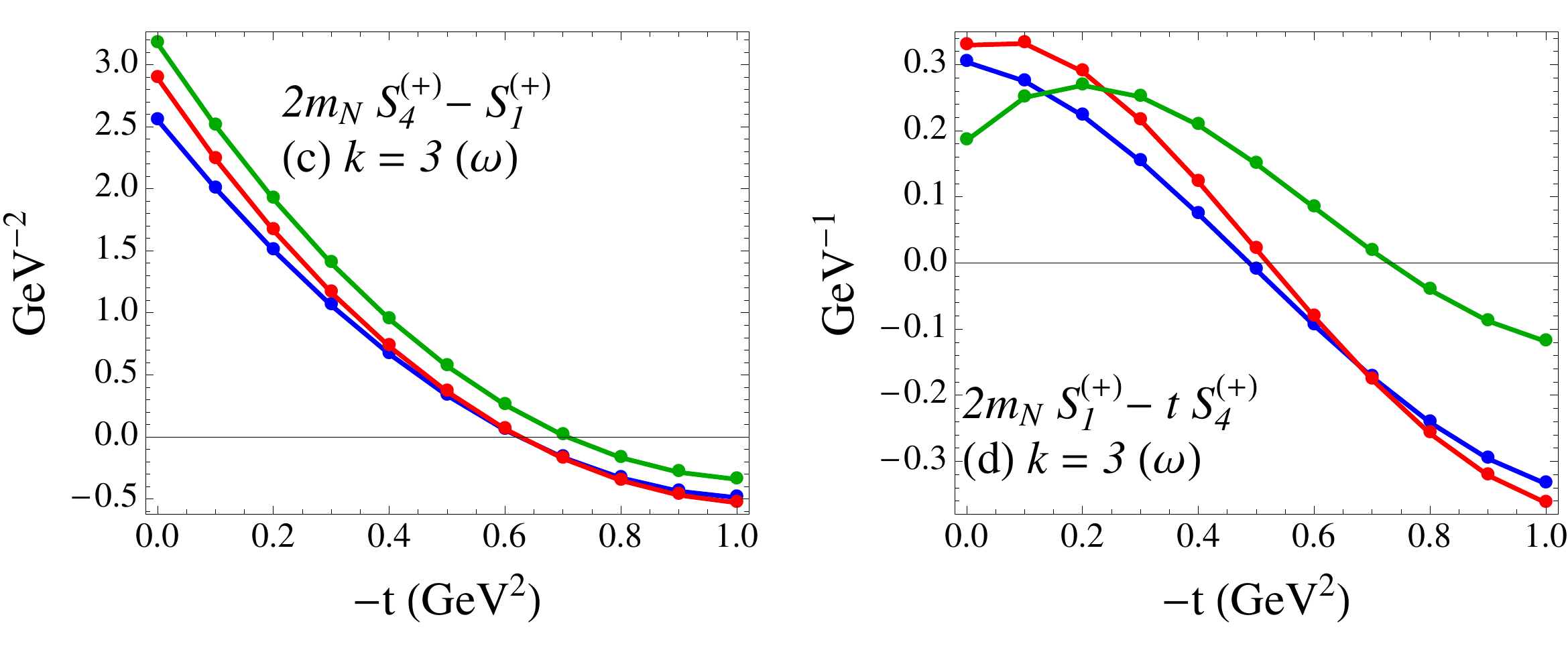}
}
\caption{\label{fig:F13Rho} Moments of the $t$-channel invariant amplitudes $F_1^{(0,+)}$ and $F_3^{(0,+)}$, Eq.~\eqref{eq:F13}, with the SAID, MAID and ANL-O models. The integral is truncated at $s_\text{max} = (2.0\gev)^2$.}
\end{figure*}

In the previous section, we exploited the relations between the scalar functions $A_i$ and the $s$-channel helicity amplitudes at leading $s$, cf. Eq.~\eqref{def:Ai}. For instance, the $t$ factor expected in the Regge residues from the factorization properties of Regge poles was readily checked using Eq.~\eqref{eq:tkin}.
However, the properties of Reggeons are best described in their rest-frame, the $t$-channel center-of-mass frame. For natural exchanges the relevant combinations are the $t$-channel natural-parity amplitudes~\cite{Mathieu:2015eia}:
\begin{subequations} \label{eq:F13}
\begin{align}
F_1 & = -A_1 + 2m_N A_4, \\ 
F_3 & = 2m_N A_1 -t A_4. 
\end{align}
\end{subequations}
$F_1$ ($F_3$) is the nucleon helicity non-flip (flip) amplitude in the $t$ channel~\cite{Mathieu:2015eia}. We now wish to compare the features of the $\rho$ and $\omega$ Regge poles obtained by FESR with other reactions sharing the same nucleon vertex. For this purpose we perform the appropriate combination of $S_i^{(\sigma)}$, from Eq.~\eqref{eq:F13}, and compare to the same quantities in $\gamma p \to \eta p$, Fig.~8 of Ref.~\cite{Nys:2016vjz}, and $\pi p \to \pi p$, Fig.~2 of Ref.~\cite{Mathieu:2015gxa}. Our results are presented in Fig.~\ref{fig:F13Rho}. We note a striking similarity between $\pi^0$ and $\eta$ meson photoproduction for the $\omega$ exchange. The moments combination for the ($t$ channel) nucleon non-flip $F_1$ displays in both cases a zero for $t \sim -0.6$ GeV$^2$. The moments combination for the ($t$-channel) nucleon flip $F_3$ displays in both cases a violation of factorization at $t=0$ and a zero for $t \sim -0.5\gev^2$. The factorization of the $\rho$ pole residues at $t=0$ is observed in both $\pi$ and $\eta$ photoproduction for the nucleon flip combination, but a zero appears for the $k=3$ moment only in $\pi^0$ photoproduction. This zero at $t \sim -0.8	\gev^2$ is shifted compared to the nucleon flip amplitudes for $\pi N$ scattering, which is at $t \sim -0.5\gev^2$. In the $\rho$ nucleon non-flip combination, the zero appears at $t \sim -0.15\gev^2$ in both $\pi^0$ photoproduction and $\pi N$ scattering. This zero was responsible for the crossover between $\pi^-p$ and $\pi^+ p$ elastic scatterings~\cite{Mathieu:2015gxa}. These similarities in the position of the zeros suggest that the zeros in the Regge residues would come from the nucleon vertex, as it is the common vertex in all these reactions.

\section{Combined fit of the FESR and observables} \label{sec:fit}
\begin{figure*}[htb]
\centerline{\includegraphics[width=\linewidth]{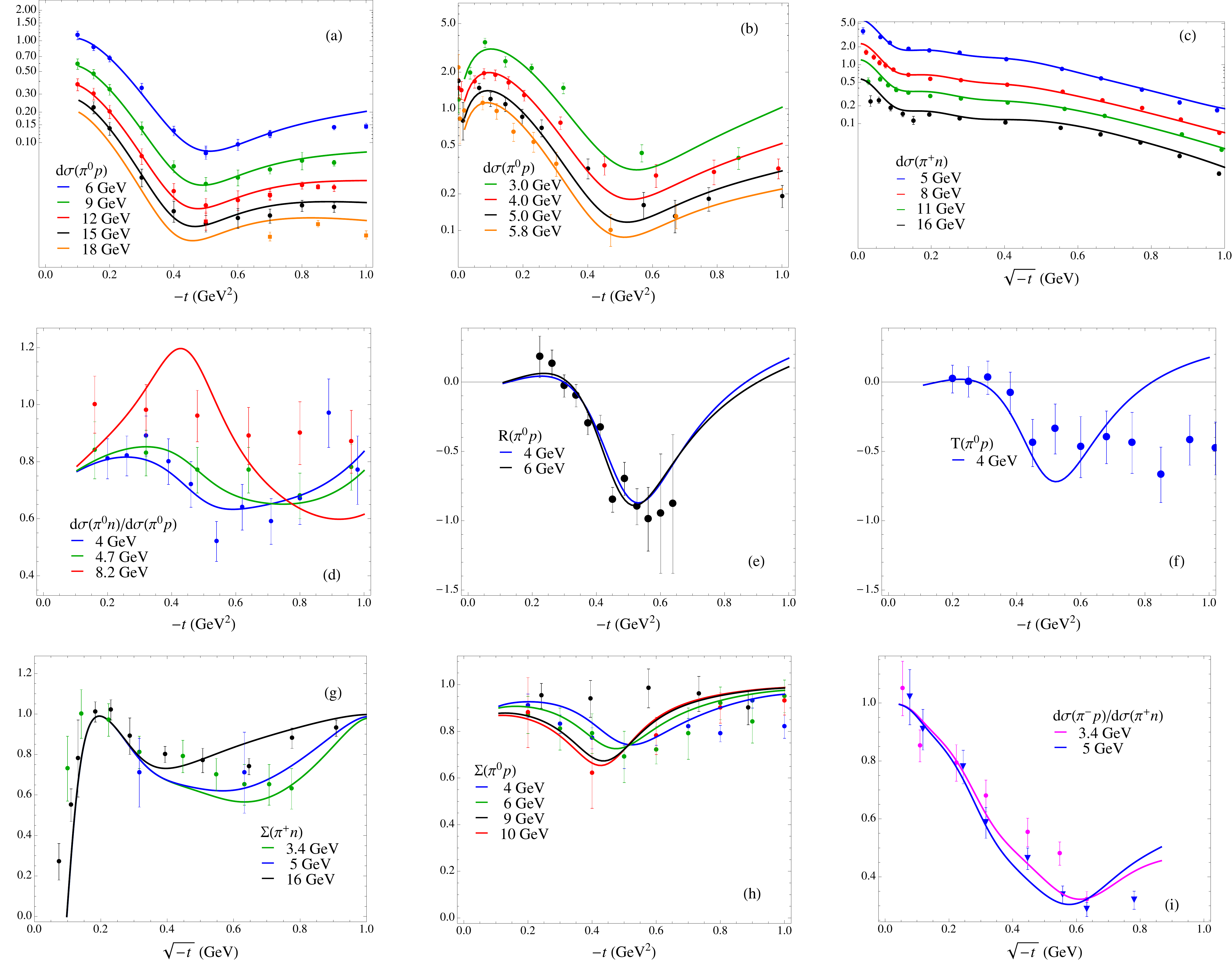}}
\caption{\label{fig:fitObs}Comparison between the observables computed with the parametrization of the amplitudes given by Eq.~\eqref{eq:imAi} and Tables~\ref{tab:solalpha}, and \ref{tab:sol} and the data from \cite{Anderson:1971xh,Barish:1974qg,Bolon:1967zy,Braunschweig:1973wd, Osborne:1973ed,Bolon:1971ac, Anderson:1970wz, AlGhoul:2017nbp, Booth:1972qp,Boyarski:1967sp,Dowd:1967zz,Heide:1968dqa,Sherden:1973fz,BarYam:1967zz,Geweniger:1969ed}. }
\end{figure*}

\begin{figure*}[htb]
\centerline{\includegraphics[width=\linewidth]{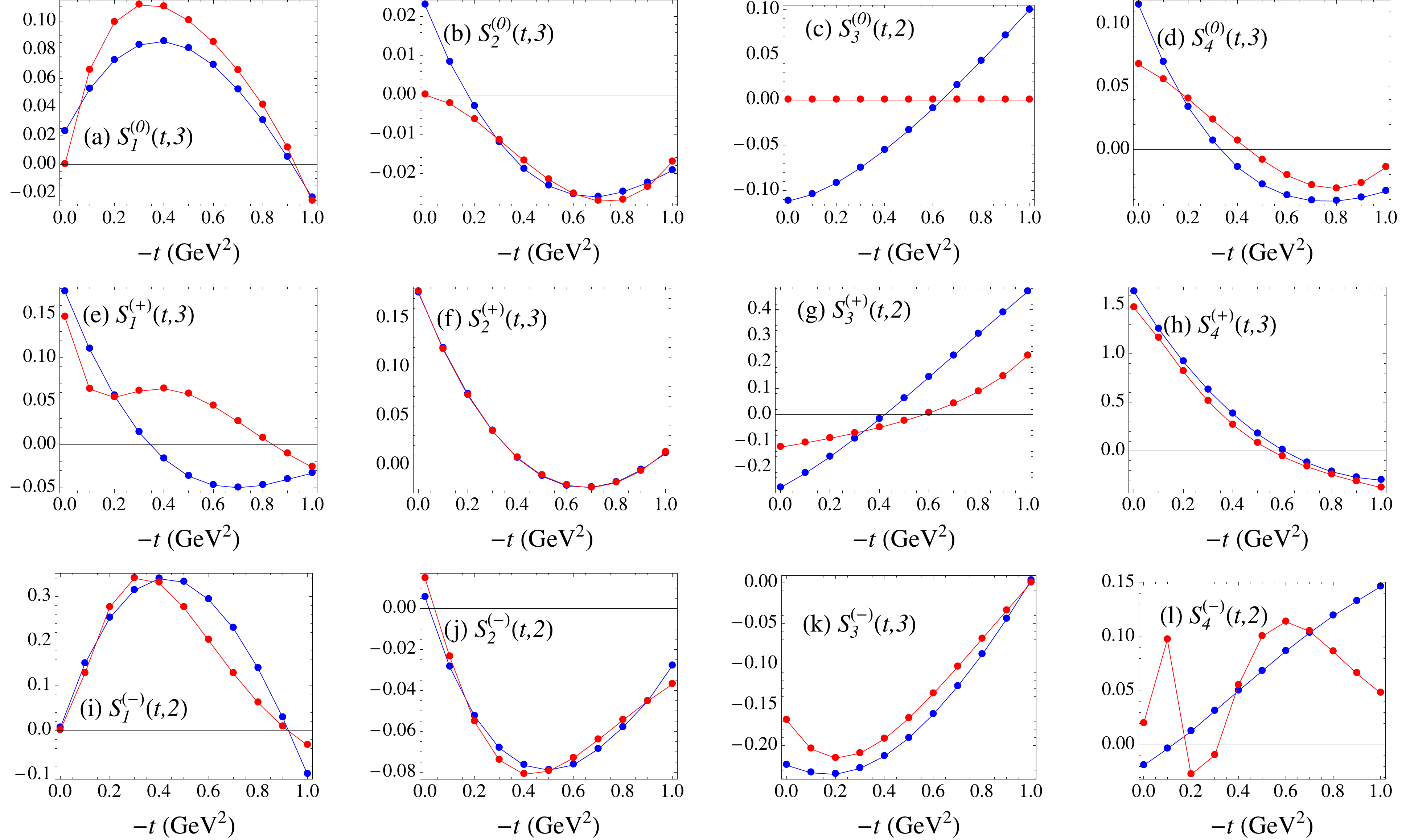}}
\caption{\label{fig:fitFESR} Comparison between the high-energy side of the FESR computed with the parametrization of the amplitudes given by Eq.~\eqref{eq:imAi} and Tables~\ref{tab:solalpha} and \ref{tab:sol} and the low energy side of the FESR using the SAID model. The cutoff is $s_\text{max} = (2 \text{ GeV})^2$.}
\end{figure*}

In the previous sections we observed the position of zeros in the moments $S_i^{(\sigma)}(t,k)$. The agreement with the expectations from Regge theory suggested the dominance of a leading Regge pole in the 12 isospin scalar amplitudes, with the possibility of subleading contributions slightly shifting  the  zeros. In this section we continue our analysis by performing a combined fit of the moments $S_i^{(\sigma)}(t,k)$ and of the high energy observables, using a Regge pole parameterization for the high-energy amplitudes. We restrict the high energy observables to the kinematical region $E_\text{lab} \ge 3\gev$ and $-t \le 1\gev^2$. In this region we have the following data sets available:
\begin{itemize}
\item [--] Differential cross section for $\gamma p \to \pi^0 p$ from Refs.~\cite{Anderson:1971xh,Barish:1974qg,Bolon:1967zy,Braunschweig:1973wd}.
\item [--] Ratio of differential cross section $\gamma n \to \pi^0 n$ over proton target from Refs.~\cite{Osborne:1973ed,Bolon:1971ac}.
\item [--] $\gamma p \to \pi^0 p$ beam~\cite{Anderson:1970wz, AlGhoul:2017nbp}, target~\cite{Booth:1972qp} 
and recoil~\cite{Deutsch:1973tg} asymmetries.
\item [--] Differential cross section for $\gamma p \to \pi^+ n$ from Refs.~\cite{Boyarski:1967sp,Dowd:1967zz,Heide:1968dqa,Sherden:1973fz}.
\item [--] Ratio of differential cross sections $\gamma n \to \pi^- p$ over $\gamma p \to \pi^+ n$ from Ref.~\cite{BarYam:1967zz}.
\item [--] $\gamma p \to \pi^+ p$ beam asymmetry from Ref.~\cite{Geweniger:1969ed}.
\end{itemize}
The observables are displayed in Fig.~\ref{fig:fitObs}.
In order to better appreciate the small $t$ region, where the pion exchange dominates, the $\gamma p \to \pi^+ n$ observables are plotted against $\sqrt{-t}$. 

We used the $S_i^{(\sigma)}(t,k)$ derived from the SAID model with the cutoff $s_{\text{max}} = (2.0\gev)^2$, computed with $k=2$ for crossing odd amplitudes or $k=3$ for crossing even amplitudes. We chose the moments $k=3$ as the moments $k=1$ didn't always present the zero pattern expected from Regge theory.  The lhs of the sum rules is evaluated at 11 points equally spaced in the range $t \in [-1,0]\gev^2$. Since we do not have any information about the uncertainties of the PWA models, and therefore of the lhs of the sum rules, we assumed an artificial constant error on each $S_i^{(\sigma)}(t,k)$, taken as the 20\% of the maximum value of each scalar amplitude.

\subsection{High-energy model}
In order to properly describe the observables and the rhs of the sum rules, our model for the imaginary part of the scalar amplitudes involves a summation of Regge pole-like terms:
\begin{align}\label{eq:imAi}
\im A^{(\sigma)}_i(\nu,t) & = \sum_j \beta^{(\sigma)}_{ij}(t) \nu^{\alpha_j(t)-1}.
\end{align}
Equating the left and right hand sides of the sum rules, this form yields,
\begin{align}
S_i^{(\sigma)}(t,k) & = \sum_j \beta^{(\sigma)}_{ij}(t) \frac{ \Lambda^{\alpha_j(t)-1}}{\alpha_j(t)+k},
\end{align}
with the cutoff given by Eq.~\eqref{eq:cutoff}.
We remind the reader that $S_2^{(\sigma)}$ stands for the sum rule evaluated with the amplitudes $A_2' = A_1 + t A_2$. In Eq.~\eqref{eq:imAi} the index $i = 2$ stands for the amplitudes $A_2^{\prime(\sigma)}$. 
In each amplitude, the summation involves one single term representing the leading Regge pole contribution. In the natural exchange amplitudes $A_1^{(\sigma)}$ and $A_4^{(\sigma)}$, we added a second Regge contribution, to have more flexibility, based on our observations from the lhs of the sum rules. The poles are the same for the same isospin components in $A_1^{(\sigma)}$ and $A_4^{(\sigma)}$, since they have the same quantum numbers. $A_1^{(\sigma)}$ and $A_4^{(\sigma)}$ are the $t$ channel nucleon helicity flip and non-flip amplitudes, respectively. There are thus 6 natural Regge trajectories: the $\rho$, $\omega$ and $a_2$,  and the $\rho$, $\omega$ and $a_2$ subleading poles, or ``daughters''.  We include only one Regge pole in the natural amplitudes $A_2^{\prime(\sigma)}$ and $A_3^{(\sigma)}$,  since the leading unnatural poles are expected to be smaller, at the same order of magnitude of a subleading natural pole.
We keep the $\pi$, $b$, $h$ and $a_1$ trajectories degenerate, and consider a $\rho_2/\omega_2$ Regge pole in the $A_3^{(0,+)}$ amplitudes. With these 2 unnatural poles, we have in total 8 trajectories, all of them linear:
\begin{align}
\alpha_j(t) = \alpha^0_j + \alpha_j^1 t.
\end{align}
The parameters of 3 natural ($\rho$, $\omega$ and $a_2$) leading trajectories and the $\pi/b/h/a_1$ trajectory are constrained around the standard values, cf.~Eq.~\eqref{eq:alpha}. The intercepts and slopes are constrained in the range $[0.3,0.7]$ and $[0.7,1.1]\gev^{-2}$, respectively. 

\begin{table}[htb]\caption{Solution of the fit for the trajectories. \label{tab:solalpha}}
\begin{ruledtabular}
\begin{tabular}{ c  | c  c  c }
& & & \\[-7pt]
$\phantom{1}j \phantom{1}$ &  $\alpha_j^0$ & $\alpha_j^1$ ($\gev^{-2}$) & role \\[3pt]
\hline && & \\[-7pt]
$1$ & 0.541 & $0.711$ & $\rho$ pole \\
$2$ & 0.316 & $0.897$ & $\omega$ pole\\
$3$ & 0.699 & $1.100$ & $a_2$ pole\\
$4$ & 0.401 & $0.661$& $\rho/\omega$ daughter \\
$5$ & -0.010 & $1.00$ & $a_2$ daughter\\
$6$ & -0.007 & $0.615$ & $\pi,b,h,a_1$ pole \\
$7$ & 1.031 & $1.770$ & $\rho_2,\omega_2$ pole\\
$8$ & 0.197 & $0.330$ & $\omega$ daughter\\
\end{tabular}
\end{ruledtabular}
\end{table}

Since all $S_i^{(\sigma)}(t,k)$ have only one extremum in the region of interest, we parametrize the residues with a second order polynomial times an exponential fall-off, 
\begin{align} \label{eq:resform}
\beta(t) & = \alpha^\kappa(t)\ t^\delta \times \beta_0 e^{bt} (1-\gamma_1 t)(1-\gamma_2 t),
\end{align}
where we omitted the indices $(\sigma)$ and $ij$.
A factor $\alpha(t)$ is needed in the $A^{(-)}_{1,4}$ and $A^{(0,+)}_{3}$ amplitudes, which involve the even signature trajectories $a_2$, $\rho_2$, and $\omega_2$. This factor cancels the unwanted ghost pole at $\alpha(t)  = 0$ that might appear in the physical region. Indeed the even signature amplitudes $A_{1,2,4}^{(-)}$ and $A_3^{(0,+)}$ have the form
\begin{align} 
A^{(\sigma)}_i(\nu,t) & = -\sum_j \beta^{(\sigma)}_{ij}(t) \nu^{\alpha_j(t)-1} \frac{1 + e^{-i\pi\alpha_j(t)}}{ \sin\pi \alpha_j(t)}
\label{eq:Aeven}
\end{align}
and have a pole at $\alpha_j(t) = 0$. Note that we did not need this factor in the $\pi$ exchange amplitudes $A_{2}^{(-)}$ since we expect the point $\alpha_\pi(t= m_\pi^2) = 0$ to lie outside the fitting region. Our fit, cf. Table~\ref{tab:solalpha}, led indeed to $\alpha_\pi\equiv \alpha_6(t) = 0$ at $\sqrt{t} = 0.107$ GeV close to the pion mass (and outside the fitting region). We thus set $\kappa = 1$ for the residues of $A^{(-)}_{1,4}$ and $A^{(0,+)}_{3}$, and $\kappa = 0$ for the others.

For completeness we quote the expression for the odd signature amplitudes, $A_{1,2,4}^{(0,+)}$ and $A_3^{(-)}$:
\begin{align} 
A^{(\sigma)}_i(\nu,t) & = -\sum_j \beta^{(\sigma)}_{ij}(t) \nu^{\alpha_j(t)-1} \frac{-1 + e^{-i\pi\alpha_j(t)}}{ \sin\pi \alpha_j(t)}.
\label{eq:Aodd}
\end{align}
The negative sign in Eqs.~\eqref{eq:Aeven} and~\eqref{eq:Aodd} is conventional. It ensures that the imaginary part of the amplitude has the same sign as the residues. 

The second factor $t^\delta$ in Eq.~\eqref{eq:resform} imposes factorization in the $A_1$ and $A_2'$ amplitudes. The poles in these amplitudes are forced to have a factorisable form with $\delta = 1$, except for the pion pole. The latter needs to have a nonzero residue at $t = 0$, in order to describe the forward peak in the differential cross section and the rapid variation of the beam asymmetry in the $\pi^+$ photoproduction.  Similarly, we do not include the factor of $t$ in the $h$ pole in $A_2^{\prime(+)}$, or in the $\omega$ leading and sub-leading poles in the amplitude $A_1^{(+)}$, as $S_2^{(+)}$  and $S_1^{(+)}$ displays a significant deviation from factorization. In the amplitude {$A_1^{(0,-)}$, the residues for both Regge contributions (the pole and the sub-leading pole) have the factor $\delta = 1$ as $S_1^{(0,-)}$ satisfy factorization at $t=0$ in good approximation.

\begin{table}[htb]\caption{Results of the fit for the residues Eq.~\eqref{eq:resform}. The factors $\beta_0$ are dimensionless. The parameters $b$, $\gamma_1$ and $\gamma_2$ are in $\gev^{-2}.$\label{tab:sol}}
\begin{ruledtabular}
\begin{tabular}{ c  c | c  c   r  r r r }
& &&&& & & \\[-7pt]
&  &  $\kappa$ & $\delta$ & $\beta_0$ & $b$ & $\gamma_1$ & $\gamma_2$ \\[3pt]
\hline &&&& & & & \\[-7pt]
$\rho$ & $\beta_{11}^{(0)}$ & $0$ &$1$ &  $\phantom{-}0.793$ & 1.806 & 0.413 & 13.08  \\
          & $\beta_{14}^{(0)}$ & $0$ &$1$ &  $-4.824$  &  0.075  & $-0.597$  & $ 0.374 $ \\[5pt]
$\omega$ & $\beta_{12}^{(+)}$ & $0$ &$0$ &  $0.744$ &   $3.131$&  $ -4.042$ & 6.876  \\
               & $\beta_{18}^{(+)}$ & $0$ &$0$ &  $-0.058$ &   $3.928$  &  $-5.514$ &   132.2 \\[5pt]
$a_2$ & $\beta_{13}^{(-)}$ & $1$ &$1$ &  $-0.099$ &   $3.624$  & $-0.028$ & 240.2  \\
               & $\beta_{15}^{(-)}$ & $1$ &$1$ &  $\phantom{-}51.91$ &   6.024 &  $-0.014$  & $-0.007$ \\[2pt]
\hline &&&& & & &  \\[-7pt]
$b$ & $\beta_{26}^{(0)}$ & $0$ &$1$ &  $0.040$  &   0.491&   $-0.870$ &  20.85  \\[5pt]
$h$ & $\beta_{26}^{(+)}$ & $0$ &$0$ &  0.881 &   0.378 &  $-2.291$  & $-1.068$ \\[5pt]
$\pi$ & $\beta_{26}^{(-)}$ & $0$ &$0$ &  0.049  &  4.557  &  5.886 & $-25.58$ \\[5pt]
\hline &&&& & & &  \\[-7pt]
$\rho_2$ & $\beta_{37}^{(0)}$ & $1$ &$0$ &  $0$  & $0$ &  0  &  $0$ \\[5pt]
$\omega_2$ & $\beta_{37}^{(+)}$ & $1$ &$0$ &  $-0.359$ &  0.035  &  0.411  &  0.385 \\[5pt]
$a_1$ & $\beta_{36}^{(-)}$ & $0$ &$0$ &  $-0.841 $ &   1.342   & $-0.999$   & $5.245$  \\[5pt]
\hline &&&& & & & \\[-7pt]
$\rho$ & $\beta_{41}^{(0)}$ & $0$ &$0$ & $-0.037$ &    0.465 &  51.644  &  2.111  \\
          & $\beta_{44}^{(0)}$ & $0$ &$0$ & 0.350 &   0.000 &   2.670 &   1.909   \\[5pt]
$\omega$ & $\beta_{42}^{(+)}$ & $0$ &$0$ & 6.896  &  3.698  & $-1.583$ &   3.623  \\
               & $\beta_{44}^{(+)}$ & $0$ &$0$ & $-0.001$ &    0.002 & $-31.57$&  $-37.13$  \\[5pt]
$a_2$ & $\beta_{43}^{(-)}$ & $1$ &$0$ & $-0.352$ &  6.776  & $22.402$ &  $-4.470$   \\
               & $\beta_{45}^{(-)}$ & $1$ &$0$ &  $-32.552$ &   7.948 &   $-2.936$  &  $-5.534$  \\[2pt]
\end{tabular}
\end{ruledtabular}\end{table}

Using the model for the residues described above, we now fit both the FESR and the observables. The observables are, in the high energy limit, 
\begin{subequations} \label{eq:HELobs}
\begin{align}
\frac{d\sigma}{dt} & = \frac{1}{32\pi} 
\left(|A_1|^2-t|A_4|^2 + |A_2'|^2 - t |A_3|^2 \right), \\
\Sigma\frac{d\sigma}{dt} & = \frac{1}{32\pi} 
\left(|A_1|^2-t|A_4|^2 - |A_2'|^2 + t |A_3|^2 \right), \\
T\frac{d\sigma}{dt} & = \frac{\sqrt{-t}}{16\pi} \im
\left(A_1 A_4^* - A'_2 A_3 \right), \\
R\frac{d\sigma}{dt} & = \frac{\sqrt{-t}}{16\pi} \im
\left(A_1 A_4^* + A'_2 A_3^* \right).
\end{align}
\end{subequations}
The model involves $18\times 4=72$ parameters for the residues and $8\times 2=16$ parameters for the trajectories. As explained in the next subsection, we choose to suppress the amplitude $A_3^{(0)}$. It reduces the total number of parameters to 84.  
The 12 $S_i^{(\sigma)}(t,k)$'s provide independent and linear constraints on the imaginary part of the scalar amplitudes. They are computed at 11 equally spaced $t$ in the region $t \in [-1,0]\gev^2$. The observables are quadratic combinations of the scalar amplitudes, which yields several local minima in the parameter space.  
We wish thus to isolate subsets of observables sensitive only to subsets of exchanges. The fit is therefore performed step by step.

\subsection{Neutral pion production}
We start by fitting the differential cross sections (on proton target and the ratio neutron over proton target), and the target and recoil asymmetries for $\pi^0$ photoproduction with only $A_{1,4}^{(0,+)}$, which are sensitive to the $\omega$ and $\rho$ exchanges. The trajectories of the leading $\rho$ and $\omega$ poles in $A_{1,4}^{(0,+)}$ are constrained around $\alpha_N(t) = 0.9(t-m_\rho^2)+1$ as specified before. An unconstrained subleading pole is added in all these amplitudes. To limit the number of parameters, we tried to use degenerate subleading trajectories in all four amplitudes $A_{1,4}^{(0,+)}$. However, such a parametrization does not result in a good description of the data. We obtain a better fit introducing a different trajectory for the subleading $\omega$ pole in the amplitude $A_{1}^{(0)}$. This is necessary to describe the target and recoil asymmetries (which would vanish with the leading $\rho$ and  $\omega$ poles only), as well as to reproduce the FESR. 

We then add the unnatural exchange amplitudes $A_{2,3}^{(0,+)}$, keeping $A_{1,4}^{(0,+)}$ fixed. We fit the $\pi^0$ beam asymmetry together with $S_{2,3}^{(0,+)}$. The $A_{2}^{\prime(0,+)}$ contains only the $b$ and $h$ poles. We impose their trajectories to be degenerate and constrained around $\alpha_U(t) = 0.7(t-m_\pi^2)$. More precisely, the intercept and slope are restricted in the intervals $[-0.2,0]$ and $[0.5,0.9]\gev^{-2}$ respectively. 

According to the SLAC measurement~\cite{Anderson:1970wz}, the beam asymmetry $\Sigma{(\pi^0 p)} \neq 1$. However the new GlueX measurement~\cite{AlGhoul:2017nbp} is compatible with $\Sigma{(\pi^0 p)} = 1$, or $A^{'(\pi^0)}_2 = A_3^{(\pi^0)} = 0 $. 
The target ($T$) and recoil ($R$) asymmetry in $\pi^0$ photoproduction are very similar. The high energy expression for these observables in Eq.~\eqref{eq:HELobs} suggest that $A_3^{(\pi^0 p)} \sim 0$. This is in contradiction with the large $S_3^{(0,+)}$ obtained from the FESR. 

The fit can describe simultaneously the asymmetries,  the cross sections and the moments $S_2^{(0,+)}$ and $S_3^{(+)}$. However, $S_3^{(0)}$ turns out to be strongly suppressed, with a large exponential suppression parameter $b$ in the residues, despite the non zero $S_3^{(0)}$. We thus choose not to include any pole in the amplitude $A_3^{(0)}$, yielding $S^{(0)}_3$  to be identical to zero.  At this stage, we do not have a resolution of this conflict between the significant moment $S_3^{(0)}$ from the low energy models and the negligible residues $\beta_3^{(0)}$ from the high energy observables.

\subsection{Charged pion production}
For the charged pion observables, we fit simultaneously the differential cross sections (on proton target and the ratio neutron over proton target), the beam asymmetry, and the $S_{1\cdots 4}^{(0,-)}$ moments. Since the pion is responsible for the forward peak in the differential cross section, we cannot separate unnatural and natural exchanges easily as we did for the neutral pion fit. We use as initial values for all the parameters related to the $\rho$ amplitudes $A_{1,4}^{(-)}$, the results obtained for the neutral pion fit. We also impose the initial condition $\gamma_1 = -30$ in the $\pi$ exchange amplitude $A_2^{\prime(-)}$. Indeed the dominance of the pion exchange in the forward direction and the charged pion beam asymmetry $\Sigma(\sqrt{-t} \sim 0.1-0.2)=1$ suggest a zero in the pion amplitude around $t \sim 0.01-0.04 = -1/\gamma_1$. We have used degenerate trajectories for the $\pi$, $a_1$, $b$ and $h$ poles, according to the expected degeneracy in Fig.~\ref{fig:trajec}. We have tried to impose the degeneracy of the $\rho_2$ pole in $A_3^{(0)}$ as well, but we obtain a better fit with a different $\rho_2$ trajectory.

The subleading $a_2$ pole in the $A_{1,4}^{(-)}$ amplitudes is necessary to reproduce the shape of the lhs of the FESR. Indeed the residue vanishes at the zero of the $a_2$ trajectory to remove the ghost pole. With only one common pole, $S_1^{(-)}$ and $S_4^{(-)}$ would have a zero at the same place and around $t = -0.63\gev^2$, the zero of the $\alpha_3(t)$ trajectory. 
With a subleading pole, we obtain a good description of $S_1^{(-)}$. The linear shape of $S_4^{(-)}$ is clearly more difficult to reproduce since the leading and subleading $a_2$ vanish at their respective zero of their trajectory. The $S_4^{(-)}$ obtained from the fitting procedure is thus the result of the polynomial dependence of the residue, trying to compensate for the NWSZ's built in the residue to give a growing moment. 

\subsection{Global fit}
Finally we performed a global fit of all neutral and charged $\pi$ observables and the FESR, keeping the parameters of the isoscalar ($\omega, \omega_2$ and $h$) and isovector negative $G$-parity ($a_2, a_1$ and $\pi$) parameters fixed, but fitting the parameters of the isovector positive $G$-parity ($\rho, \rho_2$ and $b$) that are common to both neutral and charged $\pi$ observables. The final parameters are listed in Table~\ref{tab:solalpha} (trajectories) and Table~\ref{tab:sol} (residues). The results of our model are compared to the high energy data in Fig.~\ref{fig:fitObs} and both sides of the sum rules are displayed in Fig.~\ref{fig:fitFESR}.

As expected, the $\omega$ pole trajectory $\alpha_2(t)=0.316+0.897 t$ is close to the standard result. The two trajectories in the $\rho$ amplitudes have similar intercept. The ``sub-leading'' one is thus difficult to interpret  as being a daughter or a cut. The origin of the second trajectory in the $\rho$ amplitudes is to provide enough freedom in the $t$ dependence to describe the FESR and the recoil and target asymmetry in $\pi^0$ photoproduction. Although a single $\rho$ pole would have been preferable, that was not enough to obtain a good fit. 

The sub-leading $a_2$ and $\omega$ poles have an intercept of the order of the intercept of the unnatural poles. The natural and unnatural amplitudes are then expanded to the same lowest order in the energy ${\cal O}(s^{0})$. 

\section{Conclusions} \label{sec:conclusion}
In this paper we analyzed the structure of $\pi$ photoproduction amplitudes using the Finite Energy Sum Rules (FESR). We compared the lhs of the FESR, as a function of the Mandelstam $t$, obtained from various models used in baryon spectroscopy analyses. We observed variations between the various models that could originate from different spin assignment to resonances. A different dependence on the cosine of the scattering angle in an amplitude results in a different $t$ dependence in the lhs of the FESR. Although some differences exist between the different models, we also found interesting common features. The lhs of the FESR for all 12 isospin amplitudes present at most one extremum and at most one zero for $|t|< 1\gev^2$. We discussed the possible interpretation of these zeros in Regge theory.  We also found that in all models, isoscalar amplitudes appear to violate  factorization of Regge poles residues. 

In Sec.~\ref{sec:fit} we built a flexible model allowing us to fit the FESR and the high-energy observables. Our solution involves the minimum Regge content in each amplitude: a leading Regge pole, whose trajectory is constrained around the expected values, and a second additional cut/daughter-like term in the natural exchange amplitudes. The latter allowed us to match the zero pattern in the lhs of the FESR and to describe the high-energy observables.

The solutions summarized  in Tables~\ref{tab:solalpha} and~\ref{tab:sol} can serve as a good starting point for a global fit of the experimental data in the whole energy range (from the resonances to the Regge region), together with the analyticity constraints. Once a cutoff $s_\text{max}$, moments $k$ and $t$ values have been chosen, it is straightforward to penalize the difference between the two sides of the sum rules in the fit. Another possibility is to parametrize only the imaginary part of the amplitudes, and to reconstruct the real part from the dispersion relation. However, this procedure is  more involved as it requires one to reconstruct the real part before building the observables. Hence, one must evaluate the integral for the $t$ value of each data point. The first method requires one to perform the integral only at predefined $t$ values and is therefore more suitable for fits to large data sets.

When extracting the properties of baryon resonances in the 2-3\gev region, the number of relevant partial waves grows and, with them, the number of parameters in the model. The technique we developed in this paper will certainly help to constrain this growing number of parameters. The solution we presented would be a good starting point to perform a joint fit of the low and high-energy data {\it via} the FESR, and eventually lead to a better understanding of the excited baryon spectrum. To this purpose, we made our solution available online on the JPAC website~\cite{Mathieu:2016mcy,JPACweb}. The user also has the possibility to vary the cutoff in the sum rules as well as the parameters of the high energy model, and display the resulting FESR and observables.

\acknowledgments
We thank 
M.~D\"oring, L.~Tiator, A.~Sarantsev, T.~Sato, and R.~Workman 
for useful discussions 
about the low energy models used in this study. 
This material is based upon work supported in part by the U.~S.~Department of Energy, Office of Science, 
Office of Nuclear Physics under contract DE-AC05-06OR23177 and DE-FG0287ER40365, 
National Science Foundation under Grants PHY-1415459, PHY-1205019, and PHY-1513524, 
the IU Collaborative Research Grant, the Research Foundation Flanders (FWO-Flanders),
PAPIIT-DGAPA (UNAM, Mexico) Grant No.~IA101717,
CONACYT (Mexico) Grant No.~251817 and the Deutsche Forschungsgemeinschaft (DFG).


\bibliographystyle{apsrev4-1}
\bibliography{quattro}

\end{document}